\documentclass[aps,pre,twocolumn,groupedaddress,preprintnumbers,superscriptaddress,longbibliography]{revtex4-1}

\usepackage{verbatim}
\usepackage{soul}
\usepackage{graphicx}
\usepackage{amsmath}
\usepackage{wrapfig}
\usepackage{epsfig}
\usepackage{float}
\usepackage{amsmath,amssymb,amsfonts,bm}
\usepackage{array}
\usepackage{color}
\usepackage[normalem]{ulem}
\usepackage{bbold}
\usepackage{todonotes}
\usepackage{sidecap}  

\newcommand{\ie}{i.e.\ }
\newcommand{\eg}{e.g.\ }

\newcommand{\avg}[1]{\langle #1 \rangle}

\renewcommand{\emph}{\textit}
\usepackage{xr}

\graphicspath{{./figs/}}

\begin{document}

\title{Self-organized system-size oscillation of a stochastic lattice-gas model
}

\author{Mareike Bojer}%
\altaffiliation{These authors contributed equally to this work.}
\affiliation{%
Arnold Sommerfeld Center for Theoretical Physics and Center for NanoScience, \\
Department of Physics, Ludwig-Maximilians-Universit\"at M\"unchen, Theresienstrasse 37, D--80333 M\"unchen, Germany 
}
\affiliation{%
Department of Physics, Technische Universit\"at M\"unchen, D--85748 Garching, Germany
}
\author{Isabella R. Graf}%
\altaffiliation{These authors contributed equally to this work.}
\affiliation{%
Arnold Sommerfeld Center for Theoretical Physics and Center for NanoScience, \\
Department of Physics, Ludwig-Maximilians-Universit\"at M\"unchen, Theresienstrasse 37, D--80333 M\"unchen, Germany
}
\author{Erwin Frey}%
\altaffiliation{Please send correspondence to frey@lmu.de.}
\affiliation{%
Arnold Sommerfeld Center for Theoretical Physics and Center for NanoScience, \\
Department of Physics, Ludwig-Maximilians-Universit\"at M\"unchen, Theresienstrasse 37, D--80333 M\"unchen, Germany
}
             
\begin{abstract}

The totally asymmetric simple exclusion process (TASEP) is a paradigmatic stochastic model for non-equilibrium physics, and has been successfully applied to describe active transport of molecular motors along cytoskeletal filaments. Building on this simple model, we consider a two-lane lattice-gas model that couples directed transport (TASEP) to diffusive motion in a semi-closed geometry, and simultaneously accounts for spontaneous growth and particle-induced shrinkage of the system's size.
This particular extension of the TASEP is motivated by the question of how active transport and diffusion might influence length regulation in confined systems. Surprisingly, we find that the size of our intrinsically stochastic system exhibits robust temporal patterns over a broad range of growth rates. More specifically, when particle diffusion is slow relative to the shrinkage dynamics, we observe quasi-periodic changes in length.
We provide an intuitive explanation for the occurrence of these self-organized temporal patterns, which is based on the imbalance between the diffusion and shrinkage speed in the confined geometry. Finally, we formulate an effective theory for the oscillatory regime, which explains the origin of the oscillations and correctly predicts the dependence of key quantities, as for instance the oscillation frequency, on the growth rate.

\end{abstract}

\maketitle

\section{Introduction}

Understanding collective transport phenomena is an important challenge in theoretical physics, with possible implications for biology and materials science. One-dimensional, asymmetric simple exclusion processes form a prominent class of idealized theoretical models that are amenable to detailed mathematical analyses; see for instance Ref.~\cite{Blythe2007} for a review.
Interestingly, these models appeared simultaneously in the mathematical literature as conceptual models with which to study interacting Markov processes \cite{Spitzer1970}
and in the biological literature as idealized models for ribosomes moving along mRNA during translation~\cite{MacDonald1968}; for recent reviews see Ref.~\cite{Chou2011, Appert-Rolland2015}.

The simplest version of such a model is the \textit{totally asymmetric simple exclusion process} (TASEP). 
In this one-dimensional stochastic lattice-gas model, particles move step-wise and uni-directionally from lattice site to lattice site at a constant (hopping) rate, provided that the next site is vacant.
Models of this class have been used to study the collective, directed transport of molecular motors along microtubules. 
In that context, the TASEP has been extended to include the exchange of particles between the lattice (microtubules) and the surrounding environment (cytosol) in terms of Langmuir kinetics~\cite{Lipowsky2001, Parmeggiani2003, Parmeggiani2004, Klumpp2003}.
The traffic jams predicted by these models have recently been observed experimentally~\cite{Leduc2012, Subramanian2013}, suggesting that these idealized lattice gases are indeed suitable for describing the collective dynamics of molecular motors.

In a further interesting line of research, extensions of the TASEP to dynamic lattices have been developed~\cite{Evans2007,Nowak2007a, Sugden2007, Sugden2007a, Hough2009, Schmitt2011, Reese2011, Melbinger2012, Johann2012, Erlenkaemper2012, Muhuri2013, Kuan2013, Reese2014, DeGier2014, Arita2015, Schultens2015, Sahoo2015}.
On the one hand, motivated by the transport of vesicles along microtubules that facilitate growth of fungal hyphae, or by growth of flagellar filaments, TASEP models have been considered in which a particle that reaches the end of the lattice may extend it by a single site~\cite{Evans2007,Sugden2007, Sugden2007a, Schmitt2011, Muhuri2013}. 
On the other hand, in efforts to quantify experimental observations of motor-mediated microtubule depolymerization \textit{in vitro}, dynamic lattice-gas models have proven useful for probing the regulation of microtubule length by motors that show uni-directional~\cite{Varga2006, Varga2009, Hough2009, Reese2011} or diffusive motion~\cite{Klein2005, Helenius2006, Reithmann2016}.
Recently, these models for depolymerizing molecular motors have been extended towards dynamic microtubules, in order to study the interplay between lattice growth and shrinkage~\cite{Melbinger2012, Reese2014, Johann2012, Erlenkaemper2012, Kuan2013, Arita2015}, and to understand the basic principles underlying cellular length control mechanisms~\cite{Marshall2004, Mohapatra2016}.

There are many possible extensions of these models, which are both interesting in their own right and can help us to understand important biological processes. Examples include large networks of biofilaments~\cite{Neri2011, Neri2013, Neri2013a}, limited protein resources~\cite{Lipowsky2001, Adams2008, Cook2009, Cook2009a, Brackley2012, Greulich2012, Ciandrini2014, Rank2018}, the fact that proteins in the cytosol do not form a spatially uniform reservoir because their dynamics is limited by diffusion~\cite{Lipowsky2001, Muller2005, Tsekouras2008, Tailleur2009, Evans2011, Saha2013, Graf2017}, and that proteins may be spatially confined, as they are in fungal hyphae or filopodia~\cite{Lipowsky2001, Klumpp2003, Muller2005, Pinkoviezky2014, Graf2017}. 

\begin{widetext}

\begin{figure}[t]
\centering
\includegraphics[width=\textwidth]{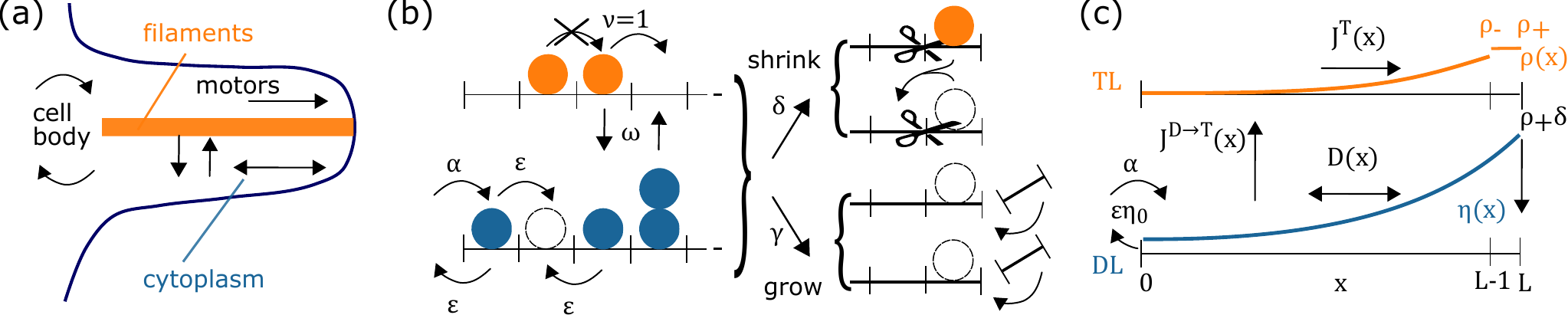}
\caption{(a) \textbf{Illustration of the dynamics in cellular protrusions.} Movements of molecular motors are indicated by black arrows, and are restricted to the cell body and the protrusion by the cell membrane. On the filament the motors move unidirectionally towards the protrusion tip, while their motion in the surrounding cytoplasm is diffusive. (b) \textbf{Illustration of the two-lane lattice-gas model.} We consider a two-lane lattice-gas model consisting of a TASEP/transport lane (TL, upper lane, occupied by orange (light gray) particles) and a diffusive lane (DL, lower lane, occupied by blue (dark gray) particles) with hopping rates $\nu\equiv 1$ and $\epsilon$, respectively. The lanes are coupled by attachment and detachment kinetics at rate $\omega$, respecting exclusion for transfer from the DL to TL. Entry and exit occurs via the first DL site only, at rates $\alpha$ and $\epsilon$, respectively. The system spontaneously grows by simultaneously appending a site to the TL and DL tip at rate $\gamma$, while both lanes shrink by a site, at rate $\delta$, if the TL tip is occupied by a particle. In the latter case, particle conservation is ensured by shifting all particles of the previous TL and DL tip site to the new DL tip site. (c) \textbf{Illustration of the particle currents and density profiles.} The density profile on the TL (DL), $\rho(x)$ ($\eta(x)$), is displayed in orange (light gray) in the upper panel (blue (dark gray) in the lower panel). The density $\rho(x)$ is discontinuous at the last TL site, with $\rho_-$ referring to the left and $\rho_+$ to the right limit. The currents (black arrows) come from entry $\alpha$, exit $\epsilon\eta_0$, diffusion $D$, attachment and detachment $J^{D\rightarrow T}$, directed movement $J^T$, and detachment due to depolymerization $\rho_{+}\delta$.}\label{fig:ModelIllustration}
\end{figure}

\end{widetext}

In this paper our goal is to study the interplay between diffusive motion and directed transport as a possible mechanism for length regulation under confinement [Fig.~\ref{fig:ModelIllustration}(a)]. This relationship is of great interest because, in contrast to diffusion, directed transport is an intrinsically non-equilibrium process. It leads to currents of motors directed towards the growing/shrinking end (tip) and so to a strong interaction between the motors and the growing/shrinking end. The combination of transport with diffusion in a semi-closed geometry has recently been studied with a conceptual model \cite{Graf2017}. This model assumes a fixed length for the system and suggests an important role for diffusion in the transport of motors to the tip. While biologically motivated exclusion in this model, and also more generally, can change the dynamics qualitatively, here we here focus on the low-density regime where exclusion only has a minor quantitative influence. Instead we extend the previous model by including length regulation. This is motivated by polymerization and depolymerization of filaments in highly dynamic cellular protrusions. For the particular choice of the growth and shrinkage dynamics, we draw our inspiration from experimental studies of microtubules, in which motor-induced depolymerization~\cite{Gupta2006, Varga2006, Varga2009, Grissom2009, Su2012} and growth by attachment of tubulin heterodimers~\cite{Mitchison1984, Howard2009} were found. Other choices such as the ``opposite" scenario where polymerization is motor-dependent and depolymerization is spontaneous, or a system with two types of motors, namely polymerizing and depolymerizing ones, are also expected to give rise to interesting phenomena but are out of the scope of the present paper.

While our motivation originates from specific biological processes, we do not want to study a particular biological system. Rather our lattice gas model (Fig.~\ref{fig:ModelIllustration}) provides us with an exemplary model to examine the combined role of diffusion and active transport for length regulation under a confined geometry.
Unexpectedly, we find that the size of our intrinsically stochastic system shows periodic behavior when diffusion is slow compared with the growth and shrinkage dynamics. This indicates that diffusion-limited transport can be an important ingredient for the occurrence of (self-organized) oscillations.

This paper is organized as follows.
In Section~\ref{sec:model} we explain the processes incorporated into the stochastic lattice-gas model and show analytical calculations for the simplest possible scenario, the stationary state, to gain a basic understanding. To check these results and explore a broader parameter regime, we continue in Section~\ref{sec:numerics} with numerical simulations. We determine the dependence of the stationary length on the growth rate and find a parameter regime in which length oscillations occur. For this oscillatory behavior we then develop an intuitive explanation. Finally, in Section~\ref{sec: Effective theory} we derive an effective theory from this intuitive explanation, and compare its predictions to the results from stochastic simulations. We conclude with a summary and discussion in Section~\ref{sec: summary}. Readers that are primarily interested in the phenomenology may want to skip the more technical part of Section~\ref{sec:mathematical_analysis}. It aims at giving a mathematical intuition about the processes constituting the presented model.

\section{Model Definition and Mathematical Analysis}
\label{sec:model}

\subsection{Stochastic Lattice-Gas Model}\label{sec: Stochastic Lattice-Gas Model}

As outlined in the Introduction, we consider a two-lane lattice-gas model in a semi-closed geometry [Fig.~\ref{fig:ModelIllustration}(b)], and extend previous work~\cite{Graf2017} by combining it with a length-regulation mechanism.
One lane, the TASEP/transport lane, TL, emulates the directed transport along filaments in cellular protrusions in terms of a totally asymmetric simple exclusion process (TASEP) \cite{MacDonald1968, Krug1991, Derrida1998}. 
It is characterized by a rate $\nu$ at which particles hop unidirectionally along the lattice, from the base towards the growing/shrinking end (tip). 
Particles exclude each other, i.e., there can be at most one particle at any lattice site and, consequently, particles can only hop forward if the site ahead of them is empty. Later we will see that exclusion is not essential for the qualitative findings discussed in this paper.
We measure all rates in units of $\nu$ and thus set $\nu\equiv 1$. 

The second lane, the diffusion lane, DL, mimics diffusive transport of motors in the cytosol [Fig.~\ref{fig:ModelIllustration}(b)], and describes it as effectively one-dimensional:
Particles perform a symmetric random walk with hopping rate $\epsilon$ to the left and right. As the density of motor proteins in the cytosol is small, we assume no particle exclusion on the DL. Hence the hopping probability is not influenced by the occupancy of the neighboring sites.

Moreover, molecular motors constantly cycle between the filaments and the surrounding cytosol by attaching to the filaments and detaching into the cytosol.
This motion is represented as follows:
At a rate $\omega$, a particle from the DL can attach to the corresponding TL site, if it is vacant, and a particle from the TL can always attach to the corresponding DL site. 

Particles can enter the system only from a reservoir via the first DL site, corresponding to motors entering the protrusion from the cell body, and similarly can only leave the system via that same site. 
Entry occurs at rate $\alpha$, and particles diffuse out at a rate equal to the hopping rate $\epsilon$. We do not model the dynamics in the cell body explicitly, as diffusion in the cell body is three-dimensional and we expect that, as a result, entry and exit events should be roughly uncorrelated. We thus approximate the cell body as an infinite reservoir.

The lanes grow by the spontaneous addition of a TL site to the TL tip at rate $\gamma$, accompanied by the simultaneous extension of the DL by one site. 
Motor-induced depolymerization is realized by cutting off the TL tip site at rate $\delta$. 
The cytoskeletal filament is considered to span the protrusion, meaning that with shrinking filament the length of the cytosolic volume shrinks as well. 
So, when the TL shrinks by one site the DL is simultaneously reduced by one site and all leftover particles, including the one responsible for the shortening event, are shifted to the new DL tip site. Thus, the DL tip site can be easily populated by several particles at once.
Since the motors can neither penetrate the membrane nor leave the system at the tip, they remain in the cytosol at the tip even when the system shrinks. 

In summary, a typical particle journey would start by the particle's entry into the system at the first DL site, followed by diffusion on this lane until it attaches to the TL and begins to hop towards the tip. Once there, it eventually cuts off the site it is occupying and joins the other particles from the previously 'lost' DL site on the new DL tip site. 
Each of these particles then diffuses on the DL until it reattaches to the TL or leaves the system at the DL's first site.

\subsection{Mathematical Analysis: Adiabatic Limit}\label{sec:mathematical_analysis}

To gain a better quantitative understanding of the system, we analyzed the stochastic dynamics of the lattice-gas model in terms of a set of Master equations, and employed a mean-field approximation to derive a set of rate equations for the density of motors on the TL and DL. 
The analysis follows Refs.~\cite{Graf2017} and \cite{Melbinger2012}, and is discussed in detail in Appendix \ref{appendix: Analytic approach}. 
Here, we will discuss the main results and their interpretation, focusing on the low-density limit and the limit of slow length change compared with particle movement, i.e.\ $\nu \equiv 1 \gg \gamma$.

To begin with, let us introduce a set of random variables to describe the state of the system: 
$L(t)$ denotes the lattice length at time $t$, configurations on the TL are indicated by a tuple of random variables $(n_i)_{i=0}^{l}$, with $n_i$ describing the occupancy of lattice site $i$. 
Each lattice site occupancy can assume the value $n_i = 1$ (occupied) or $n_i = 0$ (empty) due to mutual exclusion.
We use $l$ to denote the actual value of $L(t)$ at a specific time. 
The random variables $(m_i)_{i=0}^{l}$ representing the DL occupancy can take values in $\mathbb{N}_0$ (no exclusion).

The dynamics of the two-lane model is a difficult stochastic many-body problem, in which the bulk dynamics and the size of the system are mutually coupled.
In the limit where the bulk dynamics is much faster than the length changes, we may however assume that on the time scale over which the length of the lattice changes, the distribution of particles on the lattice is stationary \emph{(adiabatic assumption)}.
Thus we can decouple the equations for the length change and particle movement, which simplifies the mathematical analysis considerably. 
Using a mean-field approximation (see Appendix \ref{appendix: Analytic approach}) one obtains occupancy densities. We denote these as $\rho_i = \langle n_i \rangle$ and $\eta_i = \langle m_i \rangle$, where averages are ensemble averages. 

In the adiabatic limit, the stochastic dynamics of the lattice length is a simple birth-death (polymerization - depolymerization) process. Thus the system length changes as
\begin{equation}
 		\partial_t L (t) = \gamma - \delta \rho_+ ( L ) \, , \label{eq:lengthdyn}
\end{equation}
with the TL tip density denoted by $\rho_+$. Spontaneous polymerization occurs at rate $\gamma$ and motor-induced depolymerization at rate $\delta$. $L$ now  refers to the average length and is no longer a stochastic variable.
In the following, we will only consider the stationary case (and denote the stationary length by $L$). 
Thus the length change equation (\ref{eq:lengthdyn}) yields a condition on the TL tip density:
\begin{equation}\label{eq: rho+}
 		\rho_+=\frac{\gamma}{\delta} \, .
\end{equation}

In the remaining part of this Section we will formulate the current-balance equations for both lanes to derive a length-dependent expression for the particle density at the tip. Solving for the length $L$ yields the main result of this Section, Eq.~(\ref{eq:Lanalytic}). From the analysis it becomes apparent that the relevant length scale, denoted by $\lambda$ [Eq.~(\ref{eq:IntroducingLambda})], corresponds to the average distance a particle diffuses on the DL, before it attaches to the TL. Furthermore, apart from the \emph{adiabatic assumption}, meaning that the particle occupancy equilibrates fast in comparison to the length dynamics, we make use of three more approximations: first, a \emph{mean-field approximation} neglecting correlations between the occupancies at different lattice sites, justified by the low-density regime, second, the \emph{continuum limit} requiring that the number of lattice sites is large and, third, a \emph{mesoscopic limit} implying that the total attachment and detachment rate over the entire lattice are comparable to the hopping rate on the TL. 
A reader not interested in the mathematical details of the dynamics may want to skip  the remaining part of this Section.

The density profiles on the TL and DL bulk, $\rho_i$ and $\eta_i$ for $i=1, \ldots, L$, are determined by  the current balance for each lane and site [see also Fig.~\ref{fig:ModelIllustration}(c)],
\begin{subequations}\label{eq: TL and DL bulk}
\begin{align}
	\text{(TL)}\quad 0 =& + J_i^{D \to T} 
	+ (J_i^T - J_{i+1}^T)\, ,   \\
	\text{(DL)}\quad 0 
	=& - J_i^{D \to T}
	+ D_i \, , 
\end{align}
\end{subequations}
where we have defined the transport current on TL as $J_i^T := \rho_{i-1} ( 1-\rho_{i} )$, and the exchange current between TL and DL as $J_i^{D \to T} := \omega (1-\rho_i)\eta_i - \omega \rho_i$. 
Moreover, diffusion on the DL is described by $D_i := \epsilon (\eta_{i+1}-\eta_i)-\epsilon (\eta_i-\eta_{i-1})$.

At the left boundary (base of the protrusion) which is coupled to the reservoir one finds
\begin{subequations}
\begin{align}
	0 &= + J_{0}^{D \to T} 
	- J_1^T \, ,  \\
	0 &= - J_0^{D \to T} 
	+ \epsilon (\eta_{1}-\eta_0) - \epsilon\eta_0 + \alpha \, .
\end{align}	
\end{subequations}
The density current onto the TL's first site is due to particle transfer from the first site of the DL, $J^{D\rightarrow T}$ , and the transport current on the TL, $J^T$. 
For the first site of the DL there is the diffusive current onto the neighboring DL site as well as the exchange with the first site of the TL. 
Furthermore, at rate $\alpha$ particles enter the first site of the DL from the reservoir. 
This gives the corresponding influx current $\alpha$. At diffusion rate $\epsilon$ particles also exit the system from the first site of the DL, which leads to a current of $-\epsilon \eta_0$ out of the system. 

To solve these equations, we employ a continuum approximation, assuming that the lattice spacing is smaller by far than the lattice length. 
In other words, we perform a Taylor expansion in the ratio of lattice spacing $a \equiv 1$ to system size, and only keep terms up to second order. In this way, we obtain the following continuous currents with $x\in [0,L]$,
\begin{subequations}
\begin{align}
		J^T(x) &= [\rho(x)-\partial_x \rho(x)][1-\rho(x)]\, ,  \\
		J^{D\rightarrow T}(x)&= \omega [1-\rho(x)]\eta(x)-\omega \rho(x)\, ,\\
		D(x) &= \epsilon \partial_x^2\eta (x)\, ,
\end{align}
\end{subequations}
and rewrite the flux balances accordingly. 	
From the flux balances for the first sites $\eta(0),\ \partial_x \eta(0)$ and $\rho(0)$ are determined to be
\begin{subequations}\label{eq: boundary conditions}
\begin{align}
		\eta(0) &= \frac{\alpha}{\epsilon} \, ,\label{eq: boundary conditions 1}\\
		\partial_x \eta(0) &= \lambda^{-2} \, \eta(0) \, ,\label{eq: boundary conditions 2}\\
		\rho(0) &= \omega \, \eta(0) \, ,
\end{align}
\end{subequations}
having defined the length scale
\begin{equation}\label{eq:IntroducingLambda}
		\lambda \equiv \sqrt{\frac{\epsilon}{\omega}} \, .
\end{equation}
Thus the motor density at the first DL site, $\eta(0)$, equals the ratio of the particle influx rate to the particle outflux rate. $\rho(0)$ is given by the DL density at the first site from which transfer to the first TL site occurs.
The length scale $\lambda$ can be interpreted as the average distance (in units of the lattice spacing) covered by a particle on the DL by diffusion before it attaches to the TL, and it is closely related to the root mean square displacement $\propto \sqrt{\epsilon t}$ after the typical attachment time scale $t=1/\omega$. It will turn out that $\lambda$ is the intrinsic length scale of the system and most distances on the lattice will be measured with respect to this quantity.
The three boundary conditions [Eq.~(\ref{eq: boundary conditions})] will now be used as initial conditions for the bulk equations. 

First, in the low-density limit, $\rho \ll 1$, $\rho \ll \eta$, we decouple the two equations [Eq.~(\ref{eq: TL and DL bulk})]. Note that $\rho \ll 1$ implies that $(1-\rho)\approx 1$, which is equivalent to lifting the particle exclusion.
With the two initial conditions, Eqs.~(\ref{eq: boundary conditions 1}) and (\ref{eq: boundary conditions 2}), we solve the resulting second-order differential equation, $\partial_x^2 \eta (x)= \eta(x)/\lambda^2$, to give 
\begin{equation}
	\eta(x) = \eta(0) \left( \frac{1}{\lambda}\sinh\left(\frac{x}{\lambda} \right)+\cosh\left(\frac{x}{\lambda} \right)\right) \, .
\end{equation}
Sorting the bulk current balance on the TL by orders of $1/L$ implies that the TL density is the integral of the DL density that has attached to the TL,  
$\omega\int\limits_0^x \eta(y) \text{d} y =  \rho(x)$, yielding
\begin{equation}
	\rho(x)	= \rho(0) + \frac{\alpha}{\lambda}\left( \frac{1}{\lambda}\cosh\left(\frac{x}{\lambda} \right)+\sinh\left(\frac{x}{\lambda}\right) \right) \, .
\end{equation}
The resulting density profile for the low-density phase has a similar functional form as the density profile found in Ref.~\cite{Graf2017} although a static lattice was considered in that case. In particular, the exponential density increase toward the tip can be reproduced.

Regarding the last site, we expect a discontinuity in the density profile, as the hopping rules change discontinuously to accommodate growth and shrinkage. 
The left limit $\rho_-$ [see also Fig.~\ref{fig:ModelIllustration}(c)] is determined by the bulk density, while the right limit $\rho_{+}$ is fixed by the stationarity condition on the length, i.e.\ $\rho_{+}=\gamma/\delta$, Eq.~(\ref{eq: rho+}). 
The system is closed everywhere except at the first site, and consequently, the flux to the last site has to equal the flux out of the TL onto the DL, which is $\rho_{+}\delta$ to first order, 
\begin{equation}\label{eq: flux balance condition}
		J^T(L) = \rho_{+}\delta\, .
\end{equation}
This equality gives us an implicit condition on the system length $L$.

The equations for the tip dynamics become more transparent when formulated in the co-moving reference frame, as otherwise the last site is not necessarily $L$. In this frame two additional currents add to the bulk current in the previously used reference frame, the currents from relabeling due to a growth or a shrinkage event:
\begin{equation}
	J^T(x)= \rho (x)(1-\rho(x))-\gamma \rho(x)+\delta \rho_+ \rho(x) \, .
\end{equation}
Solving the flux balance (\ref{eq: flux balance condition}) yields 	
\begin{equation}
	\rho(L) = \frac{1}{2}(1-\sqrt{1- 4\gamma}) \equiv \rho_- \, , 
\end{equation}
where $\rho_- $ can be interpreted as the left limit of the density at the last site. Approximating the hyperbolic functions as exponential functions with positive argument ($\lambda \ll 1$), we obtain
\begin{equation}\label{eq:Lanalytic}
	L = \lambda \ln 
	    \left[ 
	     2 \, \frac{\lambda}{\alpha} \, 
	     \left(\rho_- - \rho (0) \right)
	     \right] \, .
\end{equation}
The higher the particle density on the TL, the faster the system depolymerizes. 
Hence, a smaller value of $\lambda$ results in a smaller steady-state length. 
This reasoning not only applies for the prefactor but also for the numerator of the argument in the logarithm. 
Here the influx into the diffusive lane ($\alpha$) is weighted by $1/\lambda$. $\rho_-$ corresponds to the critical density that depolymerizes the system at exactly the speed that is necessary in order for polymerization and depolymerization to be balanced on average. 
The bigger the critical density, the higher the stationary particle density on the TL and the longer it takes to fill the system. Thus the system has more time to grow.

\section{Numerical Analysis}
\label{sec:numerics}

So far, we derived analytical expressions for the limit of slow length change with respect to particle density equilibration. Now we want to explore the full regime, which informs us about the phenomenology of the model beyond the adiabatic regime. We therefore perform stochastic simulations of the lattice-gas model defined in Section~\ref{sec: Stochastic Lattice-Gas Model} employing Gillespie's algorithm~\cite{Gillespie1976}.
The numerical results will also be used to check the approximate analytical description for the adiabatic case, which was obtained by using a mean-field analysis.

\subsection{Choice of Parameter Space and Numerical Method}\label{sec:param_numerics}

For the numerical analysis of the system we focus on the dependence on the growth rate $\gamma$, while keeping the other parameters fixed. 
The variation of $\gamma$ causes a qualitative change in the dynamics: For small $\gamma$, the adiabatic assumption should be valid, and we expect a well-defined length, whereas for large $\gamma$ Ref.~\cite{Melbinger2012} suggests that length regulation is no longer possible \footnote{For $\gamma >1$ the system grows indefinitely as then transport on the TL is too slow to keep up with the growth dynamics.}.
We want to focus the analysis on what happens in an intermediate regime of $\gamma$: The initial length $L_0$ was set to $L_0=100$. 
We fix as attachment and detachment rate $\omega=1/L_0$, as influx rate $\alpha=0.1$, as diffusion rate $\epsilon=5.0$ \footnote{The value of $\epsilon$ corresponds to a diffusion constant that is of the order of molecular motor diffusion in cytosol.}, and as depolymerization rate $\delta=1.0$; in each case these parameters are expressed in terms of the hopping rate on the TL $\nu \equiv 1$. 
The choice of $\Omega=\omega L_0 \equiv 1$ to be of the order of the other rates is motivated by the processivity of the molecular motors, which can walk over long distances along the cytoskeletal filament before detaching \cite{Block1990}. 
It is also the theoretically interesting case as it   
guarantees that the number of attachment and detachment events over the length of the system competes with the other rates \cite{Parmeggiani2003, Parmeggiani2004}. For simplicity, we choose the same rate for attachment and detachment. However, we do not expect the qualitative results to change for different attachment and detachment rates as long as both are still taken to be small.
Moreover, $\alpha$ and $\epsilon$ together are chosen such that the density at the first site of the DL is rather small. 
As shown in Ref.~\cite{Melbinger2012},  length control in their system, that is the system neither shrinks to zero size nor grows without bound, is only feasible in the low-density parameter regime. 
Lastly, $\delta$ is chosen to be equal to the hopping rate.

We only took into account simulations where the system did not shrink to zero length but a stationary state was reached. Accordingly, we also chose the interval for the growth rate in such a way that most simulations fulfilled this criterion. The choice of $L_0$ (for fixed $\omega =\Omega/L_0$) did not influence the results in any way, as we discarded the initial behavior before the stationary state.

\subsection{Stochastic Simulations and Model Phenomenology}
\label{sec:simulations}
\subsubsection{Mean Length} 

\begin{figure}[t]
\centering
\includegraphics[width=\columnwidth]{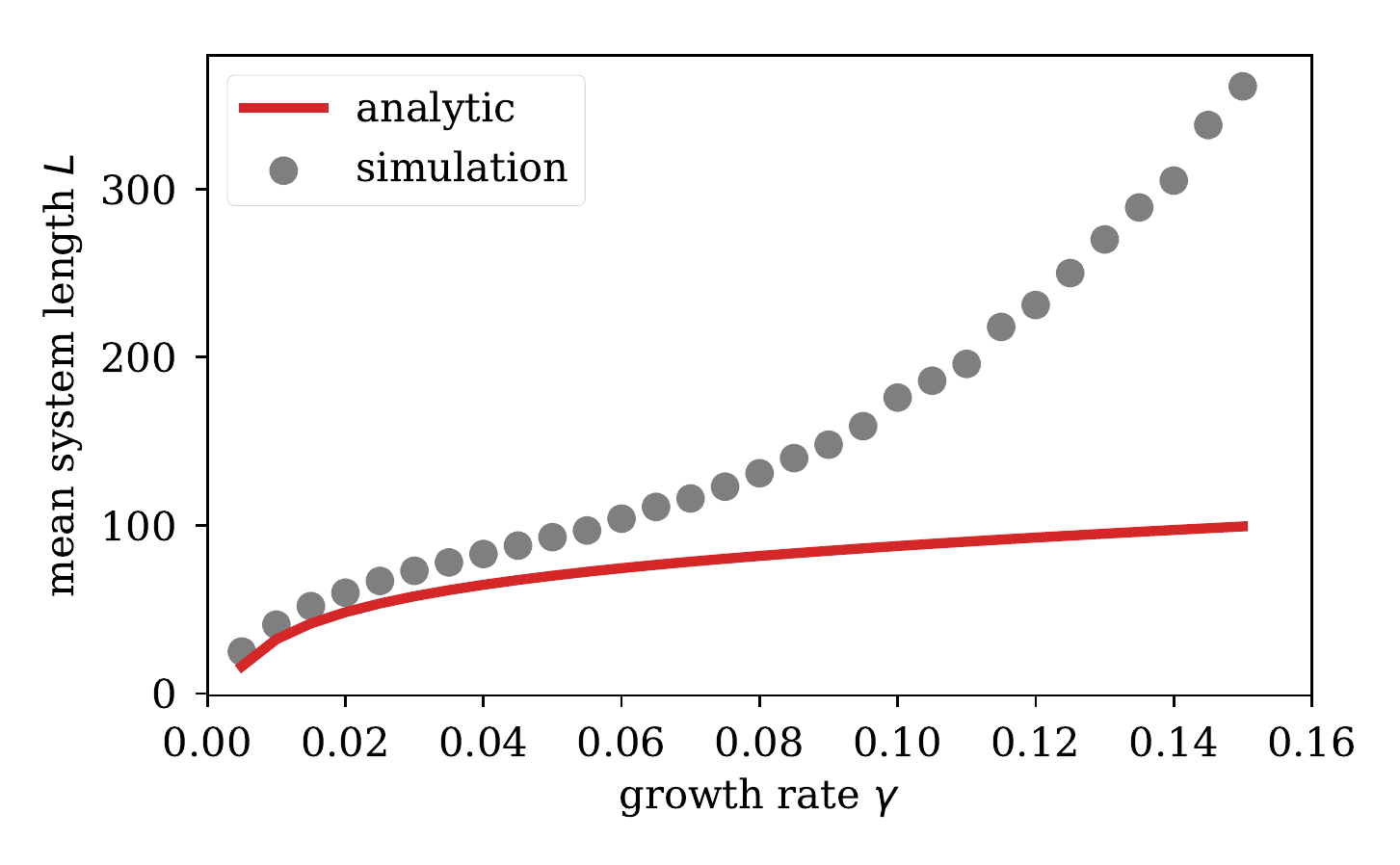}
\caption{\textbf{Mean system length}. 
Mean length of the system is plotted as a function of the growth rate $\gamma$. The analytical result (red, solid line) agrees well with the results from stochastic simulations (gray, filled circles) for small growth rates, $\gamma \ll 1$, where the adiabatic assumption is expected to hold. 
For increasing growth rates the numerical data show an inflection point at which they begin to deviate strongly from the results in the adiabatic limit. 
This indicates that the dynamics shows qualitatively new behavior for large growth rates.
}\label{fig:gammaVsL}
\end{figure}

We tested the analytical insights described in Section~\ref{sec:mathematical_analysis} by comparing them to the results of stochastic simulations.  
To begin with, we determined the mean length of the system as a function of the growth rate $\gamma$, as shown in Fig.~\ref{fig:gammaVsL}.
For small growth rates, the length increases sub-linearly with the growth rate, up to an inflection point from which it then increases super-linearly. 
As expected, the numerical result agrees nicely with the analytical results in the adiabatic limit as the growth rate tends to zero: $\gamma \to 0$ (i.e.\ when the growth and shrinkage dynamics are slow relative to the particle dynamics). However, the simulation results deviate strongly from the predictions for larger growth rates $\gamma$. 
As the adiabatic assumption was the only critical assumption in the theoretical analysis \footnote{Since we focus on the low-density limit for the TL and since there is no exclusion on the DL the mean-field approximation should be valid. }, the numerical simulations tell us that this approximation cannot be valid for larger growth rates.
On the contrary, with increasing growth rates, the particle configuration on the lattice no longer equilibrates on the time scale of the length changes. 
As a result, there must be a time lag between the length change and the equilibration of the motor configuration, and this could possibly lead to interesting dynamics. 
To explore this further, we next study the length distribution.

\subsubsection{Length Distribution}

\begin{figure}[t]
\centering
\includegraphics[width=\columnwidth]{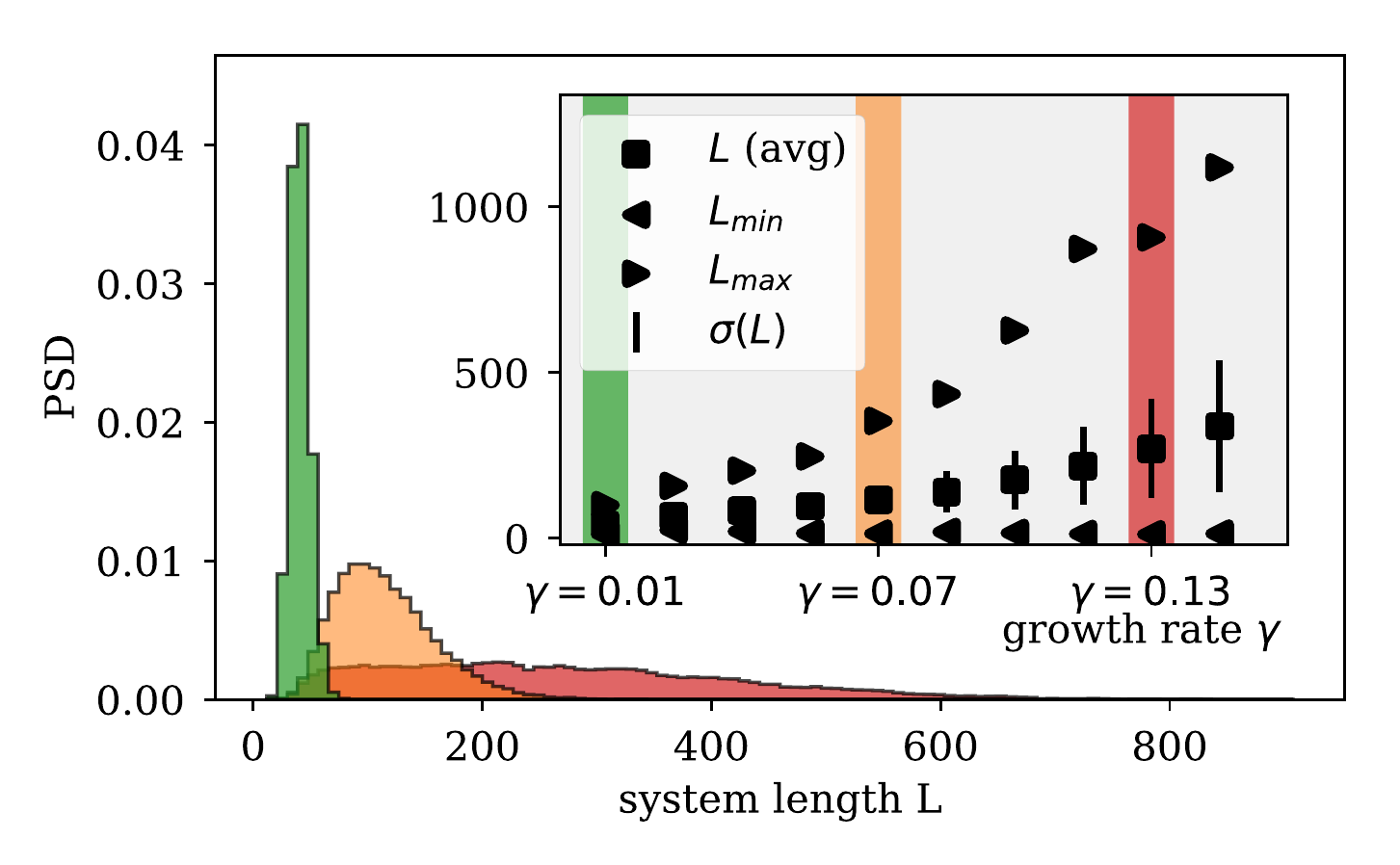}
\caption{\textbf{Length histograms} for different growth rates $\gamma$ and a simulation time of $10^7$. The larger the growth rate, the longer the average length and the broader the length distribution. The distributions are right-skewed in contrast to a Gaussian. \textit{Inset:} The average length (squares), the standard deviation (bars) of the average length, and the maximum length reached (right-pointing triangles) increase non-linearly with larger growth rates, while the minimum length attained (left-pointing triangles) remains essentially constant. The shaded areas (green (gray), orange (light gray), and red (dark gray)) correspond to the value of the growth rate $\gamma$ in the corresponding length histograms. }
	\label{fig:LHist}
\end{figure}

Figure \ref{fig:LHist} shows the length histograms for different values of the growth rate $\gamma$. In the inset, we also show the minimal and maximal lengths in comparison to the average length and the standard deviation of the length.  
We observe that for larger growth rates the length distributions become broader, while all are right-skewed. 
This right-skewness implies that we cannot approximate them as Gaussian distributions as was done in Ref.~\cite{Melbinger2012}, and so it is not feasible to use a van Kampen system-size expansion to obtain higher moments of the length distribution analytically.

From the analysis of the numerical results, we make the following observations: 
The standard deviation of the length increases with the growth rate. 
Moreover, the maximum length attained also increases with growth rate, namely faster than linearly. 
In contrast, the minimum length reached remains rather constant. 
This is surprising as, intuitively, a larger growth rate should also lead to a larger minimal length. 
Might this be connected with the suspected time lag between the length change and the equilibration of the motor configuration? 
To answer this question, we looked at a simple temporal quantity first, namely the autocorrelation function.

\subsubsection{Autocorrelation}

\begin{figure}[b]
\centering
\includegraphics[width=\columnwidth]{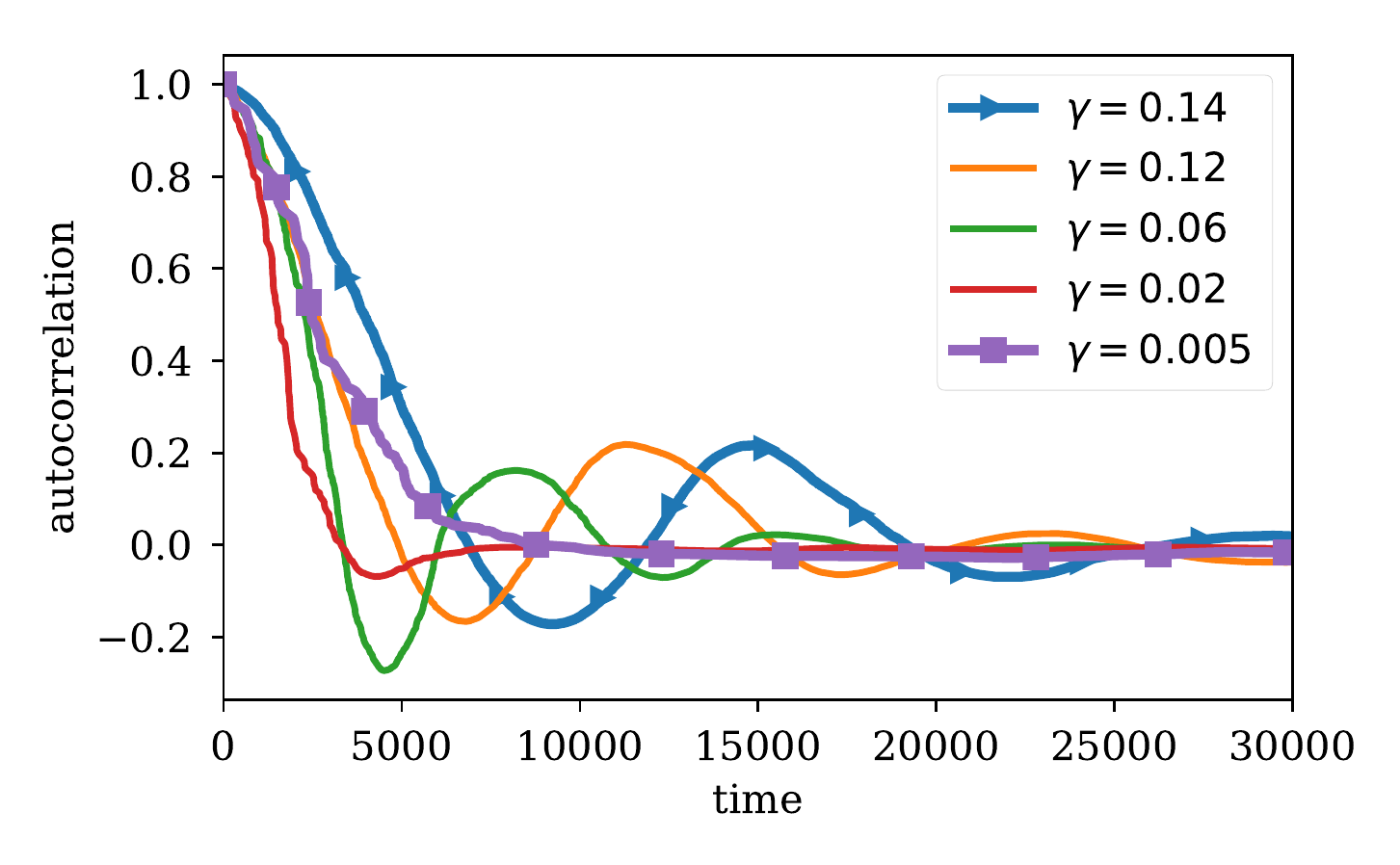}
\caption{\textbf{Autocorrelation function.} The autocorrelation functions, each of an ensemble of 1000 runs, for several growth rates $\gamma$ are compared. The autocorrelation function for the smallest growth rate $\gamma=0.005$ (purple line with squares) almost immediately decays to zero, while the autocorrelation of $\gamma = 0.14$ (blue line with triangles) oscillates with a frequency comparable to the observed length oscillations. }
\label{fig:fig5autocorr}
\end{figure}

Figure~\ref{fig:fig5autocorr} shows the ensemble autocorrelation function for different values of the growth rate $\gamma$. It can be stated in terms of the covariance between lengths at times $\tau$ and $t+\tau$, $\mathrm{Cov}(L(\tau),L(\tau+t))$, as follows
\begin{equation}
	C(t) := \avg{\mathrm{Cov}\left(L \left(\tau \right),L\left(\tau+t\right)\right)}/\sigma^2 \, \label{eq:autocorr},
\end{equation}
where $\avg{\ldots}$ denotes the ensemble average and $\sigma$ is the standard deviation of the length.

In general, we would expect the autocorrelation function $C(t)$ to decay exponentially with time, yielding an autocorrelation time that is equal to the typical internal relaxation time, i.e.\ the time scale on which a perturbation in length influences the length dynamics. 
This is indeed the case for a small growth rate ($\gamma=0.005$ in Fig.~\ref{fig:fig5autocorr}).
However, for larger growth rates, while still being enveloped by an exponential decay, the autocorrelation function oscillates with an oscillation period that increases with the growth rate. 
This indicates that for large growth rates the length is oscillating and that there might be two qualitatively different limits for the length-changing dynamics, namely for small and large growth rates, respectively. 
To study this issue further, we looked at individual time traces of the system length for small and large growth rates.

\subsubsection{Time Traces}

\begin{widetext}

\begin{figure}[t]
	\includegraphics[width=\textwidth]{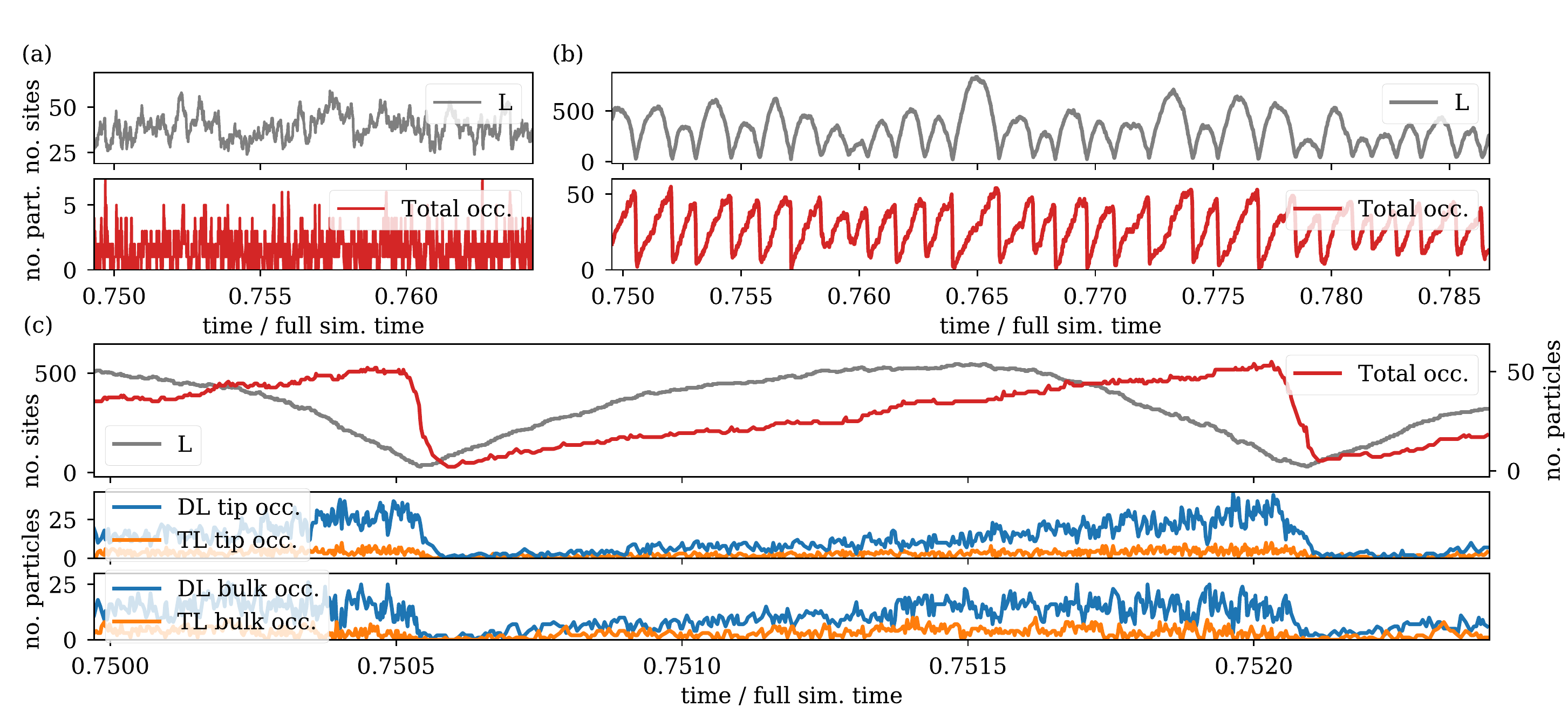}
\caption{\textbf{Time traces} for filament length, total particle number, DL occupancy, and TL occupancy for a full simulation time of $10^7$. (a) System length (gray) and total particle number (red (dark gray)) dynamics for a long time interval and small growth rate $\gamma = 0.01$: Both the length and the total particle number change stochastically. (b) System length (gray) and total particle number (red (dark gray)) dynamics for a large growth rate $\gamma = 0.14$: We observe length oscillations and a sawtooth-shaped behavior of the total particle number. (c) Zoom in for large $\gamma =0.14$: Upper panel: System length (gray) and total particle number (red (dark gray)) dynamics. Middle panel: occupancy of the TL (orange (light gray), lower line) and DL (blue (dark gray), upper line) tip neighborhood, which is chosen to consist of 20 sites from the tip. Lower panel: occupancy of the TL (orange (light gray), lower line) and DL (blue (dark gray), upper line) bulk, which corresponds to the whole lane except the tip neighborhood. We observe that the tip neighborhood is densely occupied compared with the bulk. }

\label{fig:timeSeries}
\end{figure}

\end{widetext}

Visual inspection of the time traces (Fig.~\ref{fig:timeSeries}) confirms the impression gained from the autocorrelation function that for small growth rates the length of the system fluctuates stochastically. 
In contrast, for large growth rates, the fluctuations in length are very small with respect to a dominant underlying quasi-periodic length-changing pattern, which shows roughly the same oscillation frequency as the corresponding autocorrelation function. 
This is striking, as one would not automatically assume that enhancing the \textit{spontaneous growth rate} could lead to a quasi-periodic pattern.

What might account for such behavior? 
The first question that comes to mind is whether the system is actually in stationary state and, if that were the case, how could it be reconciled with an oscillatory behavior. 
In this respect, the most obvious quantity to look at is the total number of particles that are either on the TL or on the DL. 
Is this quantity noisy or does it also show oscillatory behavior for large growth rate $\gamma$? For small $\gamma$, the total particle number behaves highly stochastically, as expected [Fig.~\ref{fig:timeSeries}(a)]. 
For large polymerization rates $\gamma$, we observe that not only the length but also the total particle number shows oscillatory behavior. 
Surprisingly, however, the time trace of the total particle number looks very different from the time trace of the system length: 
Instead of being rather symmetric within one period, the time trace for the total particle number has a sawtooth-like shape, i.e.\ the total particle number increases steadily almost during the whole period before abruptly and drastically decreasing [Fig.~\ref{fig:timeSeries}(b)]. 
Hence, the influx of particles dominates the outflux for most of the time and, in addition, the total particle number does not change synchronously with the length. 
Rather, the dynamics of the total particle number is time-delayed with respect to the length dynamics -- contrary to what one would expect if the density on the DL were more or less equilibrated. 

This suggests that the DL occupancy is far from homogeneous and that there is an intricate interaction between the motors and the length dynamics: 
From the equation of motion for the length $L$ (here considered as a stochastic variable), $\partial_t L = \gamma - \delta n_{+}$ with $n_{+}$ being the particle number at the TL tip, one expects that the instantaneous value of $n_{+}$ should be a key quantity for the length dynamics. 
It is determined by the currents along the TL and from the DL tip back to the TL tip or to the base. 
To garner information about these currents, we determined not only the total number of motors but also the number of motors located in the immediate vicinity of the tip on both the TL and DL; for specificity we chose the size of the ``tip neighborhood" to be $20$ sites.
We refer to the number of motors in the tip neighborhood and in the remaining part of the lane as  ``tip occupancy" (``tip occ."), and ``bulk occupancy" (``bulk occ."), respectively.   
These quantities are shown in Fig.~\ref{fig:timeSeries}(c) for one oscillation period. 

Based on the numerical results we can make several observations.
First, since there are typically more particles in the DL tip region of only $20$ sites than on the remaining part of the DL, the DL tip density is far higher than the DL bulk density, indicating a considerable crowding of particles at the tip.
Secondly, the DL tip occupancy in particular increases over almost the whole oscillation period before drastically decreasing only at the very end (similarly to the total particle number). 
Hence, although the system is already shrinking, the DL tip density continues to increase. 
This suggests that there is no communication between the DL tip density and the reservoir throughout most of the shrinkage phase: as diffusion is finite, there is no instantaneous equilibration between the (higher) density at the tip and the reservoir density. 
Only when the system is already very short, the cluster at the tip is released into the reservoir. 

\begin{figure}[t]
\centering
\includegraphics[width=\columnwidth]{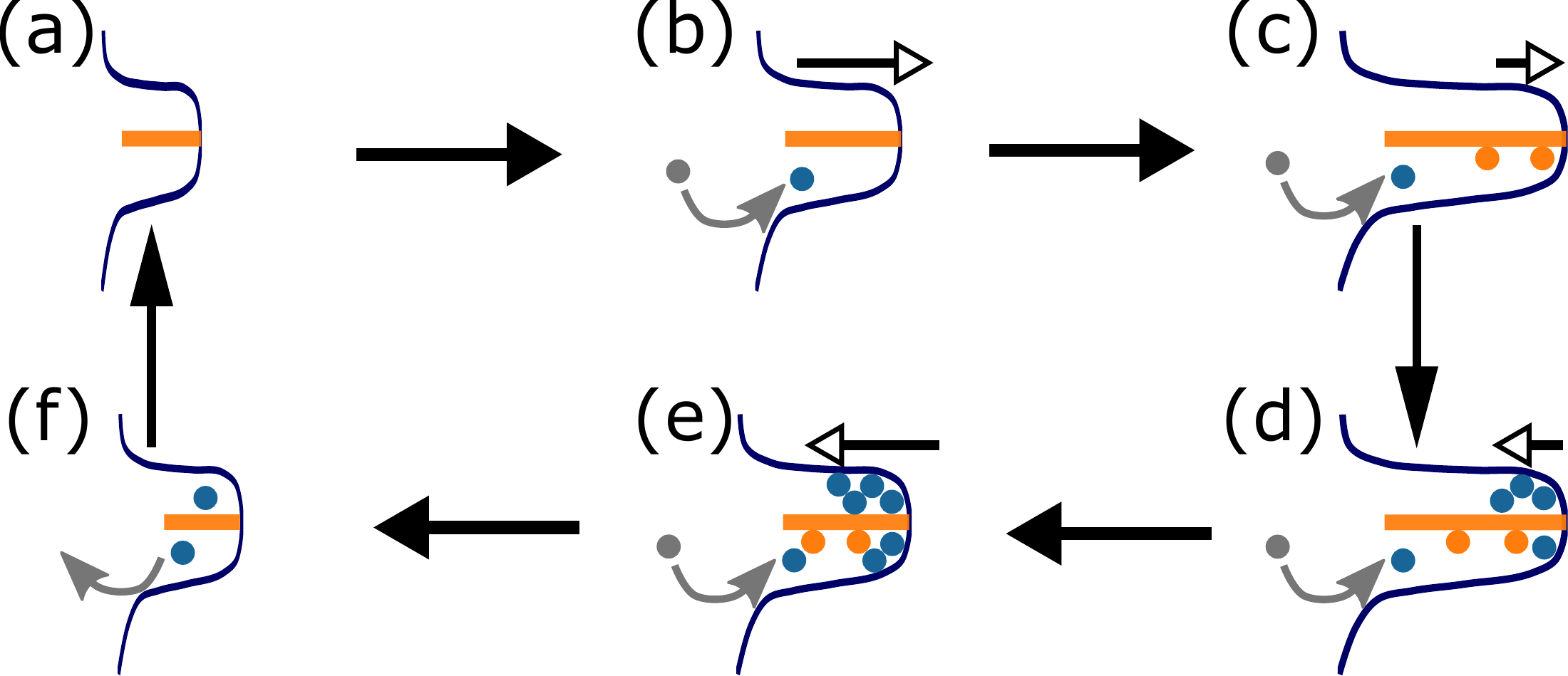}
\caption{\textbf{Intuitive picture for the occurrence of length oscillations.} 
Starting from a short and empty system (a), the only two processes possible are growth and influx of particles from the reservoir into the system (b). Once attached to the TL particles start walking towards the tip, away from the reservoir, and the system grows while new particles enter (c). Since growth is slow compared with transport of particles on the TL, the particles on the TL ``catch up" with the tip. Furthermore, due to the finite diffusion and the closure at the tip, the particles then begin crowding at the tip, turning the growth phase into a shrinkage phase (d). During the shrinkage phase more and more particles accumulate at the tip as new particles still enter from the reservoir on the left while the system shrinks from the right (e). Only when the system has become very short, is diffusion of particles fast enough that particles which accumulate at the tip can leave the system by exiting into the reservoir (f), leaving behind a short and empty system (a), from which the next oscillation cycle can begin anew.
}\label{fig:intuitive_picture}
\end{figure}

This suggests the following mechanism (see Fig.~\ref{fig:intuitive_picture}): 
Diffusion of the particles is slow relative to shrinkage, so that as shrinkage proceeds the particles cluster more and more at the tip and do not come into contact with the reservoir at the left end [Fig.~\ref{fig:intuitive_picture}(e)]. 
Hence, they cannot leave the system as long as its length is not yet sufficiently short for diffusion to be competitive. 
Only when the length of the system falls below a critical value [Fig.~\ref{fig:intuitive_picture}(f)], can the motors diffuse fast enough to reach the first site of the DL and get out of the system. 
This then happens quickly, as the reservoir particle density is very low and many motors have accumulated at the DL tip that all exit the system at around the same time, equilibrating the DL tip density with the reservoir density.
Following this reasoning, this critical length should then depend on the diffusion rate $\epsilon$ together with the effective shrinkage speed, as these two parameters determine the typical length that the particles can move away from the tip before the system further shrinks. 
If the system then becomes depleted of particles [Fig.~\ref{fig:intuitive_picture}(a)], there are no more particles at the TL tip and, as shrinkage is assumed to be particle-induced, the system can only grow. 
Since even for ``large" growth rates $\gamma$, growth is considerably slower than the TL hopping rate, $\gamma \ll 1$, particles begin to move toward the tip as the system grows [Fig.~\ref{fig:intuitive_picture}(b)] and finally reach the tip and accumulate there [Fig.~\ref{fig:intuitive_picture}(c)], turning the growth phase into a shrinkage phase (particles ``catch up" with the TL tip) [Fig.~\ref{fig:intuitive_picture}(d)]. 

Notably, this mechanism, which is based on the particle accumulation at the DL tip, heavily relies on the particle conservation, since particles can leave the tip region only via the diffusive lane. 
If this were not the case, particles could simply leave the tip region via an exit rate, effectively reducing the clustering at the tip and so shortening the extended shrinkage phase.

\section{Effective theory for the oscillatory regime}
\label{sec: Effective theory}
So far, we have built up a heuristic mechanism from an analysis of the numerical data. To examine the validity of the suggested heuristic mechanism, and to gain a more quantitative understanding of the oscillations in the parameter regime considered, we now construct an effective, semi-phenomenological theory. The effective theory incorporates the main ideas of the heuristic picture, and we will check how closely its predictions fit the numerical results.

\begin{figure}[t]
\begin{center}
\includegraphics[width=0.5\textwidth]{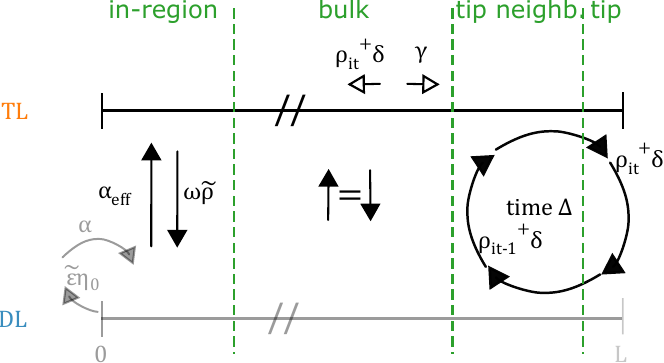}
\end{center}
\caption{\textbf{Schematic of the effective model.} We split the system into four regions, and use an effective description of the DL, restricting our analysis to the TL. In the in-region we have attachment at an effective in-rate $\alpha_{\mathrm{eff}}$ and detachment at rate $\omega$. The bulk region links the in-region to the tip neighborhood and we assume that in the bulk attachment and detachment balance. For the tip neighborhood we assume that particles which have detached at the tip a time $1/\omega$ before reattach to TL in the tip neighborhood and need another time $\Delta - 1/\omega$ to reach the TL tip again, yielding a recursion relation for the tip density $\rho^{+}_{\mathrm{it}}$. Finally, at the tip we have detachment at rate $\delta \rho^{+}$, and the corresponding shrinkage of the system, and spontaneous growth. The tip and the tip neighborhood are described in the co-moving frame.}
\label{fig:schematic_effective}
\end{figure}

The theory is based on an effective description of the diffusion lane, and on the idea that, depending on where particles detach from the TL, they either reattach to it after an average time $1/\omega$ (which is the inverse of the attachment rate), or leave the system. We first divide our system qualitatively into four regions (see Fig.~\ref{fig:schematic_effective}). From base to tip, these are the ``in-region", the ``bulk", the ``tip neighborhood" and the ``tip":
\begin{itemize}
\item The in-region is close to the base: Here, newly entered particles attach to the TL (via DL), and detach from the TL at rate $\omega$.
\item The tip: The last site on the TL at which growth and shrinkage (together with detachment of the triggering particle) occur. 
\item The tip neighborhood: Here, particles that have previously detached from the tip reattach to the TL. We neglect detachment and further reattachment, as we assume that particles which detach in the tip neighborhood reattach in the same region, balancing each other out.
\item The bulk: This merely serves as a linker region between the ``in-region" close to the base and the ``tip neighborhood" close to the tip. Here, we assume that attachment and detachment of particles balance each other out (particles that detach there  also reattach there).
\end{itemize}

In summary, we assume that particles that enter the system, and do not immediately leave it again, attach to the TL in the in-region. They then either detach there again and return to the reservoir, or they walk on the TL towards the tip. Furthermore, particles that detach at the tip reattach to the TL in the tip neighborhood after an average time $1/\omega$. Moreover, growth and shrinkage occur at the tip. 

Note that the division of the system into those regions is motivated by key components of the system dynamics such as the coupling to the reservoir at the base, the particle dynamics on and between the lanes and the length-changing dynamics at the tip.
It is however a theoretical construct and instead of fixed boundaries there will be continuous transitions between the different regimes in the real system.

As we have seen in Section~\ref{sec:simulations}, the total number of particles in the system increases almost throughout the oscillation period, including the greater part of the shrinkage phase. As a first step, we determine the effective rate at which particles enter the system, and then attach to the TL. This rate will not equal the ``bare" in-rate $\alpha$, as particles can also leave the system again before attaching.

What is the probability, $\mathrm{Prob} (\mathrm{leaving})$, that a particle that has just entered leaves the system again before attaching to the TL? To answer this question we assume that the length of the system is considerably larger than the length of a typical journey of a particle on the DL before it attaches to the TL, and discuss the influence of a short length separately below. By carefully keeping track of all possible exit paths we determine $\mathrm{Prob} (\mathrm{leaving})$ as
\begin{align*}
\mathrm{Prob} (\mathrm{leaving}) =  1- \frac{\sqrt{\omega \epsilon}}{\tilde{\epsilon}} +  \mathcal{O} (\omega)\, 
\end{align*}
(see Appendix, \ref{appendix:effective_in_rate}), where we allow the exit rate from the system, $\tilde{\epsilon}$, to be different from the diffusion rate $\epsilon$. As a result, the effective on-rate onto the TL is given by
\begin{align}
\alpha_{\mathrm{eff}} \approx \frac{\alpha }{\tilde{\epsilon}} \sqrt{\omega \epsilon}
\label{eq:effalpha}
\end{align}
to lowest order in $\omega$. It is proportional to the ratio of particle influx $\alpha$ to particle outflux $\tilde{\epsilon}$ from and back into the reservoir itself, which can be interpreted as the density in the reservoir. Furthermore, the effective on-rate onto the TL increases with the attachment rate $\omega$, as expected, and with the diffusion rate $\epsilon$, since for a higher diffusion rate (compared with the exit rate $\tilde{\epsilon}$) particles diffuse further into the system. Using this effective entrance rate we now proceed to our effective TASEP model. 

First, we estimate the length of the in-region $l_I$, since - due to attachment and detachment here - its length influences the density. To do so we model a typical particle on the DL (which does not leave the system immediately) until it attaches to the TL, as a symmetric random walker with reflecting boundary at $x=0$. Attachment to the TL follows a Poisson process at rate  $\omega$. Assuming that the particle starts at $x=0$, and diffuses with diffusion constant $\epsilon$ (lattice spacing 1), we find that the average lattice site until which the particle has diffused when attaching to the TL is given by $\langle x \rangle \pm \sigma (x) = \sqrt{\epsilon/\omega} \pm \sqrt{\epsilon/\omega} = \lambda \pm \lambda$ (see Appendix, \ref{appendix:effective_in_region}). Here $\lambda = \sqrt{\epsilon/\omega}$ is a characteristic length scale of the system (see Section~\ref{sec:mathematical_analysis}). 

Since we assume the in-region to extend from the base into the system, we will approximate it as the region  $[0, 2 \lambda]$ on symmetry grounds. The left (right) boundary corresponds to the average distance a particle travels before attaching to the TL ($\lambda$) minus (plus) the standard deviation (also $\lambda$). The length of the in-region is determined by the characteristic length scale $\lambda$:
\begin{align}
l_I \approx 2 \lambda \, .
\end{align}

With this, we now determine the density profile in the in-region, which we assume to equilibrate quickly on the time scale of the oscillations.  
Furthermore, we assume that attachment is evenly distributed over the whole in-region, yielding an attachment rate of $\alpha_{\mathrm{eff}}/l_I$ per site. In the low-density and continuum limit, together with the hopping transport on the TASEP and detachment of particles at rate $\omega$, this yields a density profile in the in-region $\tilde{\rho} (x) = \alpha_{\mathrm{eff}}\left[1- e^{- \omega x} \right]/ \left( l_I \omega \right)$, $x \in [0,l_I]$ (see Appendix \ref{appendix:effective_in_region}). In particular, the density at the right end of the in-region is given by
\begin{align}
\tilde{\rho} (l_I) = \frac{\alpha}{2 \tilde{\epsilon}} \left[ 1-e^{- 2 \sqrt{\epsilon \omega}}\right]\, .
\label{rholI}
\end{align}
It increases with the density in the reservoir, $\alpha/\tilde{\epsilon}$, and also with both the diffusion rate $\epsilon$ and the attachment and detachment rate $\omega$. Note that we measure time in units of the hopping rate $\nu \equiv 1$ on the TASEP, and length in units of the lattice spacing $a \equiv 1$. 

We introduced the bulk region in order to interpolate between the densities in the in-region and in the tip neighborhood. Since the bulk region is sufficiently far from the reservoir and from the tip (at least when the length of the system $L \geq 4 \lambda$) we assume that attachment and detachment approximately balance, and so the density is approximately constant and equal to $\tilde{\rho} (l_I)$ [Eq.~(\ref{rholI})]. 

For the analysis of the dynamics in the tip neighborhood and at the tip, we switch to a different reference frame, namely starting at the tip and reaching into the tip neighborhood, co-moving with the tip. The tip neighborhood represents that part of the system within which particles that have detached from the TL tip typically diffuse on the DL before reattaching to the TL. Thus, we assume that the tip neighborhood has the same length as the in-region $l_T= l_I= 2 \lambda$ as for both the average distance traversed before attaching to the TL is essential. We will now substantiate the idea that particles that have detached from the tip reattach back to the TL: We suppose that particles that detach at the tip reattach to the TL on average after time $1/\omega$. Furthermore, they then walk to the tip during an additional average time $l_T/2 = \lambda$, since on average they attach to the TL at a distance $l_T/2$ away from the tip and take one directed step during time $1/\nu = 1$. So, the tip density at time $t$, $\rho^{+} (t)$, influences the tip density at time $t+1/\omega + \lambda \equiv t+ \Delta$, $\rho^{+} (t+\Delta)$.
We determine $\rho^{+} (t+\Delta)$ as the steady-state of the dynamics in the tip neighborhood and at the tip that results from the usual TASEP dynamics in the low-density and continuum limit combined with growth and shrinkage, and attachment at rate $\delta \rho^{+} (t)/l_T$ per site in the tip neighborhood (see Appendix \ref{appendix:effective_recursion_relation}). 

In summary, we imagine that particles that enter the tip region start ``cycling" there: They detach at the tip, diffuse in the tip neighborhood, reattach to the TL, walk back to the tip, detach again and so on (Fig.~\ref{fig:schematic_effective}). As long as $\epsilon < 1/\omega$, the average distance $\lambda$ to the tip after reattaching to the TL is less than the average walking distance $1/\omega$ on the TL, so most particles that reattach to the TL reach the tip. 

This procedure yields a recursion relation for the tip densities $\rho^{+}_{\mathrm{it}}$ at times $t_{\mathrm{it}} = \mathrm{it} \times \Delta$ (see Appendix \ref{appendix:effective_recursion_relation} for an explicit formula) that could, in principle, be used to determine the time evolution of the tip density. So far, however, we have implicitly assumed that the length of the system, $l$, is infinitely long, $l \gg \lambda$, and we have not considered how the physics changes for comparatively short system lengths. In particular, we ignored the fact that the shorter the system, the less likely particles that have previously detached from the tip are to reattach to the TL, as they may now leave the system beforehand. So, there will be some minimal length at which the majority of particles that had previously been in the tip region has left the system. From about this point the system starts growing again.

To estimate this minimal length, we consider a 1D system with injection of particles (=detachment) at rate $r$ at site $l$ (tip), symmetric diffusion at rate $\epsilon$ within the system, outflux (=reattachment) of particles at rate $\omega$ everywhere, and an additional outflux of particles at rate $\tilde{\epsilon}$ at site $0$. In the steady-state and with a continuum approximation, the reattachment probability of a particle detaching at the tip at length $l$ can be approximated as
\begin{align}
p_{\mathrm{reattach}} (l) \approx 1- F \exp \left( -\frac{l}{\lambda} \right)
\label{eq:reattach}
\end{align}
for $l \gg \lambda$ (see Appendix, \ref{appendix:effective_minimal_length}), where 
\begin{align}
F = \frac{2 \lambda^3}{\varphi+2 \varphi \lambda +(1+\varphi) \lambda^2 + \lambda^3} \, .
\label{eq:F}
\end{align}
Here, $\varphi = \epsilon/\tilde{\epsilon}$ is the ratio between the diffusion rate and the exit rate.
As expected, the reattachment probability decreases with decreasing length $l$, and has the characteristic length scale $\lambda$. Furthermore, $F$ increases with decreasing $\varphi$, and so the reattachment probability decreases with decreasing $\varphi$. As a result, for a larger exit rate compared with the diffusion rate (small $\varphi$), the reattachment probability is small. 

We have chosen the time interval $\Delta$ in such a way that during time $\Delta$ a given particle that has detached at the tip diffuses in the DL, reattaches and walks back on the TL to the tip. So, in order for a particle to remain in the system, it needs to reattach to the TL each time it has detached and so, it needs to reattach back for all lengths $l_{\mathrm{it}}$ the system attains at times $t_{\mathrm{it}} = \mathrm{it} \times \Delta$. We have $p_{\mathrm{survival}} \left(\{l_{\mathrm{it}} \}_{\mathrm{it}=1, \ldots, n} \right) = \prod_{\mathrm{it}=1}^n p_{\mathrm{reattach}} (l_{\mathrm{it}})$ after a series of lengths $\{ l_{\mathrm{it}}\}_{\mathrm{it}=1, \ldots, n}$. We further define the minimal length as the length where approximately $50\%$ of the particles that were in the system at maximal length have left it.  Making the rough assumption that the system shrinks at a constant velocity $v \approx \gamma/2$, which is half the maximal growth speed, we find 
\begin{align}
l_{\mathrm{min}} \approx \lambda \ln \left[ \frac{2 \lambda F}{\gamma \Delta \ln (2)} \right]\, ,
\label{eq:lmin}
\end{align}
with $F$ as defined before, Eq.~(\ref{eq:F}) (see Appendix \ref{appendix:effective_minimal_length}). This means that, to leading order, the minimal length is determined by the typical length scale $\lambda$. The weak logarithmic dependency on the inverse growth rate $1/\gamma$ arises from the fact that the growth (and shrinkage) speed scales with $\gamma$.

Taking these considerations together, we find the following recursion relation for the tip densities $\rho^{+}_{\mathrm{it}}$ and the lengths $l_{\mathrm{it}}$ at times $t_{\mathrm{it}} = \mathrm{it} \times \Delta$: 
\begin{widetext}
\begin{subequations}
\label{eq:soltiplengthredboth}
\begin{align}
\rho^{+}_{\mathrm{it}} &=\left\{\left[\rho^{+}_{\mathrm{it}-1}\delta^2 \;  \mathbb{1}\! \left(l_{\mathrm{it}-1} {-} l_{\mathrm{min}} \right)-A\right] + \sqrt{\left[\rho^{+}_{\mathrm{it}-1}\delta^2 \; \mathbb{1}\!\left(l_{\mathrm{it}-1} {-} l_{\mathrm{min}} \right) - A\right]^2 + 2 B \left[\rho^{+}_{\mathrm{it}-1}\delta \; \mathbb{1}\!\left(l_{\mathrm{it}-1} {-} l_{\mathrm{min}} \right)+C\right]}\right\} \Big/ B \label{eq:soltiplengthreda} \, ,\\
l_{\mathrm{it}} &= l_{\mathrm{it}-1} + \gamma \Delta -  \rho^{+}_{\mathrm{it}-1}\delta \Delta \, ,
\label{eq:soltiplengthred}
\end{align}
\end{subequations}
\end{widetext}
with initial condition $\rho^{+}_0=0$ and $l_0=0$ (which, however, does not influence the long-term behavior, as in the case of the stochastic simulation). Furthermore, 
\begin{align}
A&= \delta \left(1 - \gamma \right) + \gamma \left(1-\gamma \right) - \delta \tilde{\rho} (l_I) \left(2-\gamma \right) \, , \\
B&=2 \delta \left[\delta (1-\tilde{\rho}(l_I)) + \gamma \right] \, ,  \\
C&=(1-\gamma) \tilde{\rho} (l_I) \, , 
\end{align}
where we use $\tilde{\rho} (l_I)$, $l_{\mathrm{min}}$ and $\Delta$ as defined before. Eq.~(\ref{eq:soltiplengthred}) derives from the growth and shrinkage dynamics (constant growth at rate $\gamma$ and motor-induced shrinkage at rate $\delta \rho^{+}$) during the time interval $\Delta$, and $\mathbb{1}$ denotes the Heaviside step function. 

\begin{figure}[t]
	\includegraphics[width=\columnwidth]{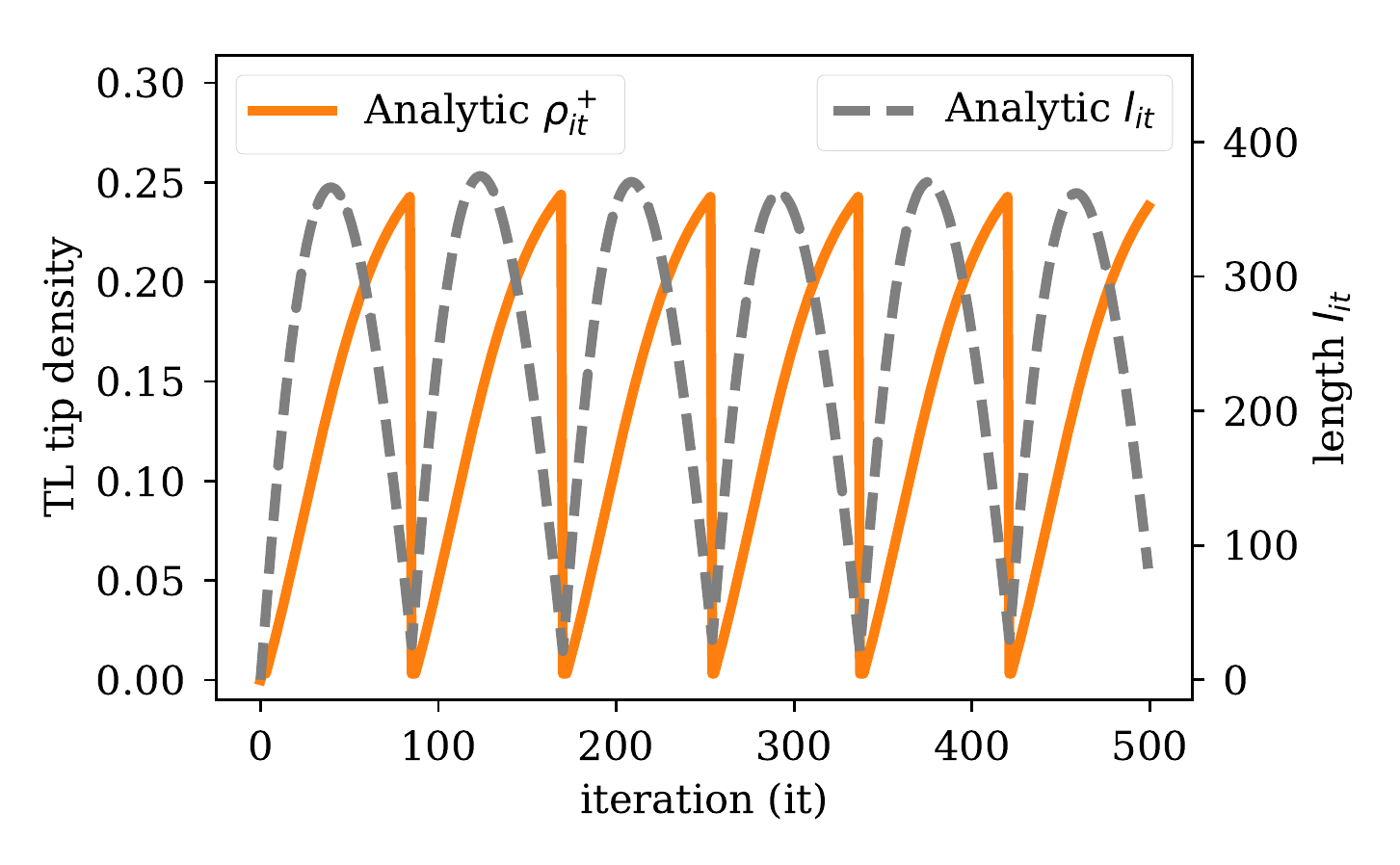}
	\caption{\textbf{Solution of the recursion relation} for the tip density (orange, solid line) and the length (gray, dashed line) as a function of the iteration step, for $\gamma=0.14$. Both show periodic behavior, but while the growth and shrinkage phases are rather symmetric for the length dynamics, the tip density exhibits a sawtooth shape.}
	\label{fig:AnalyticTtraces}
\end{figure}

Solving this recursion relation numerically, we now compare the predictions of our effective theory to the outcomes of simulations. To begin with, let us look at the result of the recursion relation, Eq.~(\ref{eq:soltiplengthredboth}), itself, which is shown in Fig.~\ref{fig:AnalyticTtraces}. In line with the stochastic simulations [Fig.~\ref{fig:timeSeries}(c)], the length changes periodically with relatively symmetrical growth and shrinkage phases, while oscillations of the tip density, in contrast, follow a sawtooth pattern. 

\begin{figure}[t]
\includegraphics[width=\columnwidth]{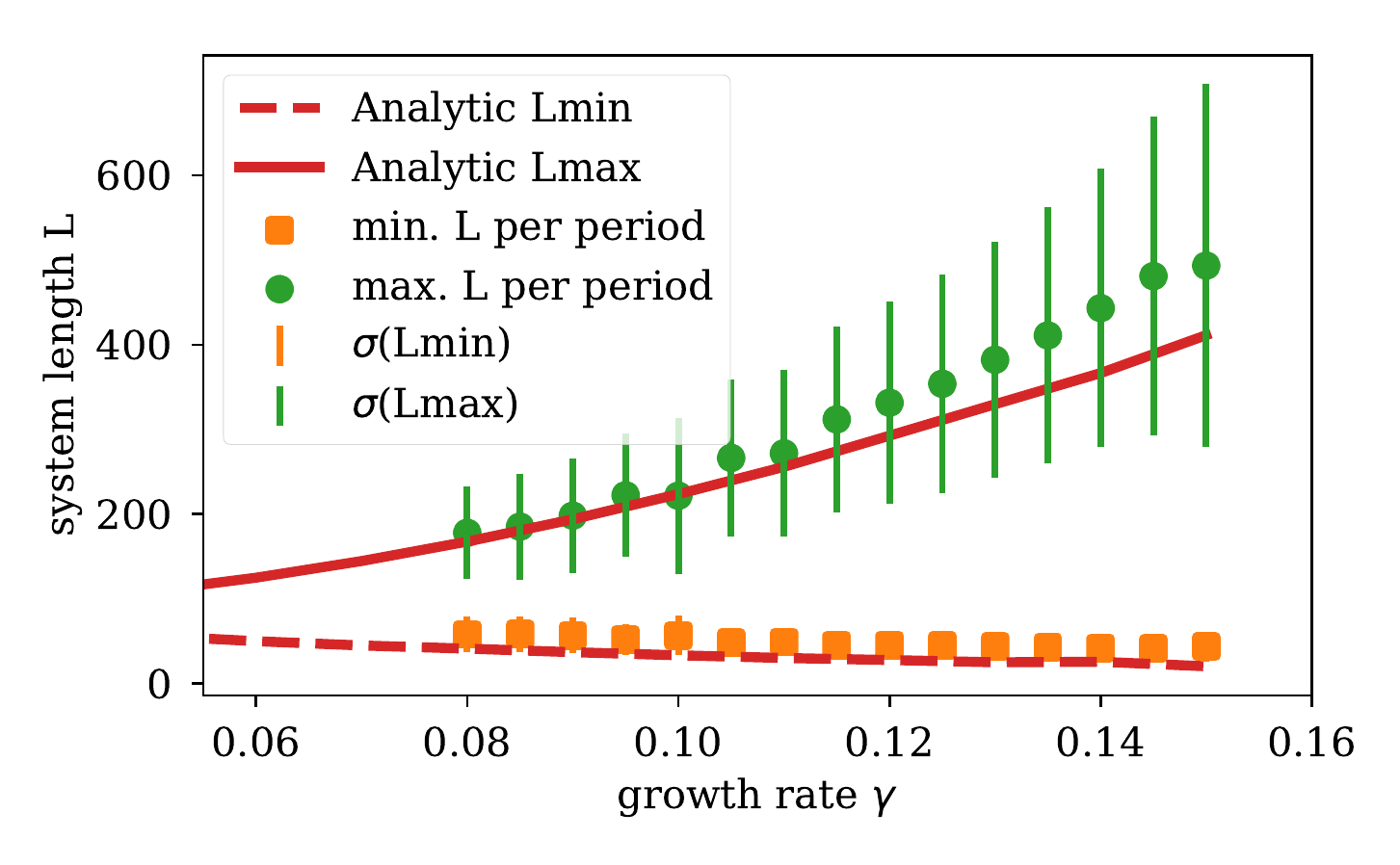}
\caption{\textbf{Minimal and maximal length.} The average minimal (maximal) length per oscillation period from stochastic simulations is compared with the prediction from the effective theory. From the stochastic simulations we determined the minimal and maximal length for each oscillation period, and the average minimal (maximal) length is depicted with orange squares (green circles), with error bars representing the corresponding standard deviation. Note that the average minimal length is approximately independent of the growth rate, in contrast to the average maximal length. The prediction from the effective theory is shown with red lines: As in the stochastic simulations the maximal length (solid line) increases with the growth rate $\gamma$, whereas the minimal length (dashed line) is only weakly dependent on the growth rate, decreasing slightly with increasing growth rate.}
\label{fig:oscillationsMinMax}
\end{figure}

For a more quantitative comparison, we have numerically determined several quantities from the recursion relation and compared them to the results from simulations. 
In accordance with the stochastic simulation, we find that the minimal length is largely independent of the growth rate $\gamma$ (Fig.~\ref{fig:oscillationsMinMax}) with a tiny decrease in minimal length for increasing growth rate in both stochastic simulations and the analytic prediction. This is what we would expect, as the turning point from shrinkage to growth should mainly be determined by the point at which diffusion (rate $\epsilon$) is fast enough relative to the shrinkage dynamics to enable the tip cluster to equilibrate with the reservoir, and thus  the system to quickly deplete. 

Second, not only the turning points from shrinkage to growth, but also the inflection points from growth to shrinkage are important. In numerical simulations, not only the maximally reached length during the full simulation (see again Fig.~\ref{fig:LHist} for more details) but also the average maximal length of the system per oscillation period increases faster than linearly with the growth rate (Fig.~\ref{fig:oscillationsMinMax}). This behavior is reproduced by our effective theory insofar as it also exhibits a faster than linear increase in the  maximal length per oscillation period with the growth rate $\gamma$ over the parameter range considered. Comparing the prediction of the effective theory with the average maximal length per period from simulations, we find quite good quantitative agreement.

\begin{figure}[t]
	\includegraphics[width=\columnwidth]{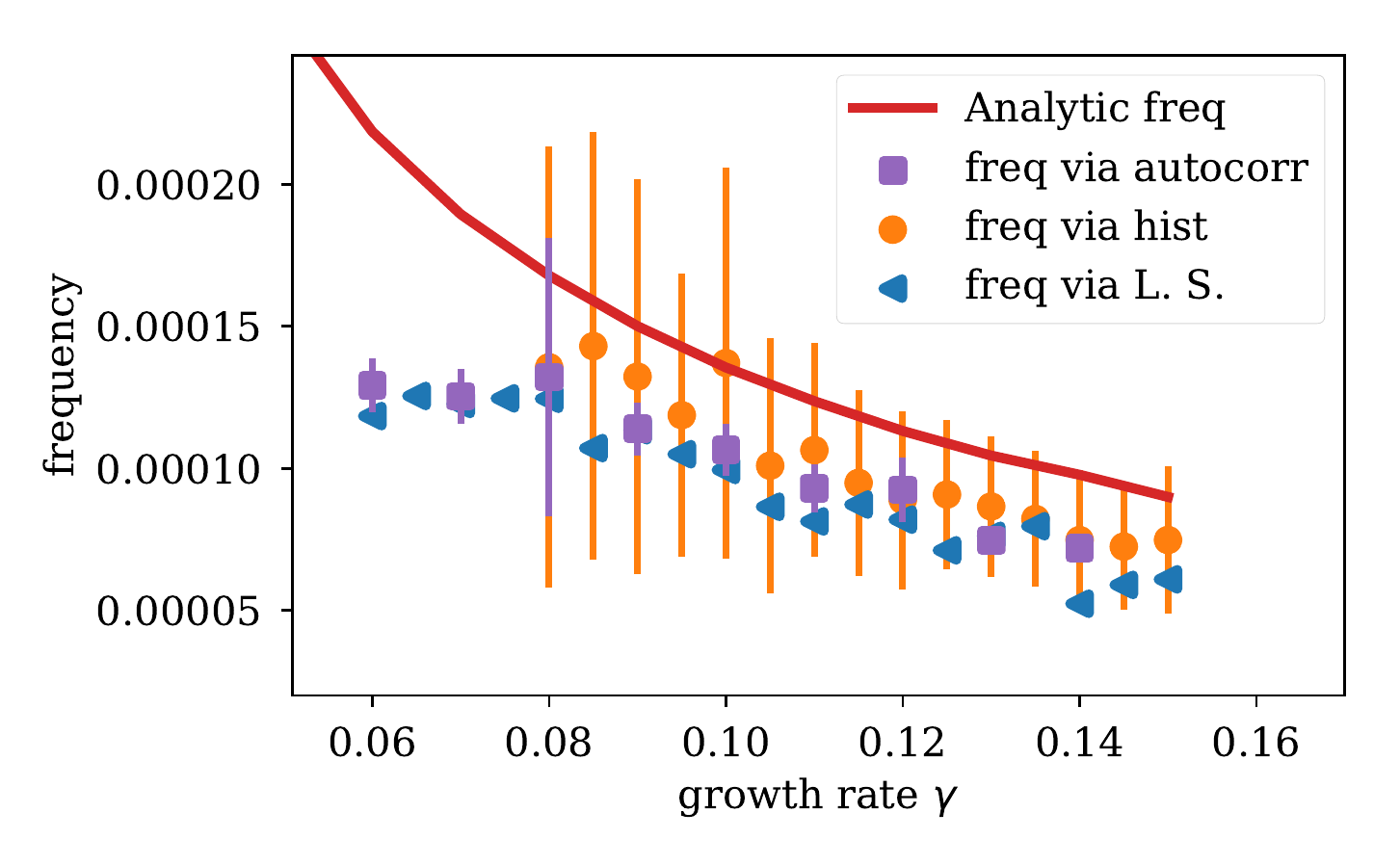}
	\caption{\textbf{Oscillation frequency.} The oscillation frequency from stochastic simulations is compared with the prediction from the effective theory. We determined the length oscillation frequency from the stochastic simulations, first, by determining the autocorrelation oscillation frequency (freq via autocorr, purple squares), second, by evaluating the distribution of duration between two adjacent minima (freq via hist, orange circles) and, third, by performing a Lomb Scargle analysis comprising a sine fit (freq via L.\ S., blue left-pointing triangles). For the distribution approach, where a bound on the maximal frequency is used as explained in the Appendix \ref{appendix: Oscillatory behaviour}, the distribution average and standard deviation (error bars) are depicted.  For the Lomb Scargle analysis the most probable frequency is shown. Clearly, for all methods the oscillation frequency decreases with larger growth rates. For very small rates noise masks the oscillation, such that the methods employed cannot determine the frequency correctly. Thus, results from the stochastic simulations are only shown for growth rates $\geq 0.06$. 
The predicted oscillation frequency from the effective theory is shown as a solid red line. It displays the same qualitative behavior as the result from stochastic simulations. }
	\label{fig:oscillationsFreqs}
\end{figure}

Apart from its amplitude (difference between maximal and minimal length), the oscillation is also characterized by its frequency.  
Only with the suggested intuitive mechanism in mind, it is not clear a priori how the frequency should depend on the growth rate $\gamma$: 
There are two possible, opposing mechanisms. On the one hand, growth (and shrinkage \footnote{The critical density given by Eq.~(\ref{eq: rho+}) increases linearly with the growth rate, so for larger $\gamma$ the shrinkage speed should increase as well.}) increase with larger growth rate $\gamma$, so the oscillation period (frequency) should decrease (increase) with growth rate $\gamma$. On the other hand, for larger growth rate, the amplitude increases as well, namely faster than linearly, and so, the oscillation period (frequency) should increase (decrease). Furthermore, it is not clear how fluctuations in length influence the oscillation frequency. In summary, it is difficult to predict from the intuitive picture alone how the oscillation frequency depends on the growth rate.

We can, however, use our effective theory and the recursion relation to numerically determine the ``analytical" oscillation frequency. We find that the analytical oscillation frequency decreases with increasing growth rate $\gamma$ (Fig.~\ref{fig:oscillationsFreqs}). As mentioned above, visual inspection of the autocorrelation functions already suggests that the same is true for the stochastic simulations, and this is confirmed by different methods to determine the oscillation frequency from the stochastic simulations (Fig.~\ref{fig:oscillationsFreqs}): In both the simulation results and the analytical prediction, the oscillation frequency at $\gamma=0.14$  is around half of its value at $\gamma=0.08$.  Note that for smaller growth rate $\gamma$ it is very hard to determine an oscillation frequency from the stochastic simulations as the oscillation is largely obscured by stochastic noise.

All in all, in the parameter regime considered, our effective theory agrees nicely with the results from stochastic simulations (Sec.~\ref{sec:numerics}), supporting the intuitive picture on which the effective theory is built.

\section{Summary and Discussion}
\label{sec: summary}

In summary, we have studied a semi-closed system consisting of two coupled lanes, a TASEP lane and a diffusive lane which, at the tip, spontaneously grow, and shrink when a particle reaches the tip of the TASEP lane. We find two qualitatively different regimes for small and large growth rates, respectively, which differ in the dynamics of length change: For small growth rates, length change is mainly stochastic, while for large growth rates oscillatory patterns dominate. 

The occurrence of those oscillatory patterns relies on the accumulation (crowding) of particles at the dynamic tip during the shrinkage phase [Fig.~\ref{fig:intuitive_picture}(d)]. This crowding leads to a positive feedback mechanism for shrinking [Fig.~\ref{fig:intuitive_picture}(e)], as each particle that reaches the TASEP lane tip further shrinks the system. The crowding is resolved only after a time delay, namely when the system size becomes comparable to the finite diffusion length. Then exchange of particles can occur between the tip region and the reservoir at the base, and the tip density equilibrates with the reservoir density [Fig.~\ref{fig:intuitive_picture}(f)], finally turning the shrinkage phase into a growth phase [Fig.~\ref{fig:intuitive_picture}(a)]. As transport on the TASEP lane is fast compared with the growth of the system, particles entering the system from the reservoir [Fig.~\ref{fig:intuitive_picture}(b)] ``catch up" with the growing tip, and start accumulating there [Fig.~\ref{fig:intuitive_picture}(c)]. As soon as the crowding reaches a critical value, the whole process begins over again. 

We provide a deeper quantitative understanding of the length oscillations by formulating an effective theory. It relies on the intuitive explanation we propose for the occurrence of the oscillations, namely cumulative crowding of motors at the tip due to finite diffusion, and correctly predicts the dependence of the oscillation frequency and amplitude on the growth rate, validating our intuitive picture. 

From this intuitive picture it is evident that the emergence of the periodic behavior crucially depends on the finite diffusion speed, which - together with particle confinement - enables crowding of particles. To our knowledge, oscillatory patterns have not been observed in any similar lattice-gas model. We attribute this to the fact that in those models diffusion had not been taken into account explicitly, or only in terms of a homogeneous reservoir, corresponding to infinitely fast diffusion.

In our system, in the limit of infinitely fast diffusion, the equilibration between the DL tip and the reservoir takes place infinitely fast, and the density on the DL is homogeneous. So, in this limit our model reduces to the model discussed in Ref.~\cite{Melbinger2012}.

On a broader perspective, the time delay due to a finite diffusion speed in a confined geometry also seems to be crucial for the occurrence of oscillatory behavior in other systems, such as in recent models for the Par or Pom protein systems~\cite{Walter2017, Bergeler2018} and for mass-conserving reaction-diffusion systems~\cite{Halatek2018}. In general, delay times have been associated with periodic behavior in well-mixed systems as well~\cite{Vilfan2005, Novak2008a}. Based on our analysis, we believe that it would be interesting to further explore how time delays can emerge intrinsically in a spatially extended non-equilibrium system, and under what conditions this leads to robust oscillations.

\begin{acknowledgments}
We thank Silke Bergeler, Matthias Rank, Emanuel Reith\-mann, and Patrick Wilke for critical reading of this manuscript and for helpful comments. 
This research was supported by the German Excellence Initiative via the program ``NanoSystems Initiative Munich'' (NIM). MB and IRG are supported by a DFG fellowship through the Graduate School of Quantitative Biosciences Munich (QBM). MB and IRG contributed equally to this work.
\end{acknowledgments}

\cleardoublepage

\appendix

\newcommand{\p}[1]{\left(#1 \right)}
\renewcommand{\d}{\text{d}}
\renewcommand{\emph}{\textit}

\section{Analytic approach}\label{appendix: Analytic approach}
In the following we perform in detail the calculations leading to the steady state density profiles as sketched in Sec.~\ref{sec: Stochastic Lattice-Gas Model}. In particular, we will elaborate on the used approximations, i.e.\ the adiabatic assumption, the mean field approximation, the continuum limit and the mesoscopic limit. We start with some comments on the used notation. \\
We denote the first, \ie the leftmost, site by ``$0$" and the last site by ``$L$". Indices will be used to denote site numbers. Moreover, the results below are stated in terms of $\rho_{j(t)}(t)$, \ie using the index at time $t$, not $t+\text{d} t$. This is necessary to clarify as site indices change due to length changes. Occupancy numbers $n$ will be approximated by occupancy densities $\rho$ ($\eta$) on the TL (DL). Often we simply denote $\avg{L}$ by ``$L$". ``$\avg{}$" represents the ensemble average.\\
We begin with the adiabatic assumption which allows us to decouple length change and particle dynamics. We perform the argument exemplarily for the TL. The first step is to write down the probability for a certain lattice site to be occupied. 

Any tuple of length $l$ with entries zero (=empty) or one (=occupied) describes a possible state of the TL with length $l$, \eg $\left( n_0 =1, n_1 =0, n_2 = \dots ,\ \dots\ , n_l=1 \right)$. Let us denote the complete set of such tuples as $\Omega(l)$, and the number of elements it contains by $|\Omega(l)|$.  $(n_i)^l_{i=0, j}$ describes the $j$-th element of this set. 
In this notation the probability of a site $i$ to be occupied with one particle at time $t+\d t$ can be written as
\begin{widetext}
\begin{equation}\label{eq:exactoccupationnumberdynamics}
\text{P}\p{n_{i(t+\d t)}(t+\d t)=1}=
\sum\limits_{l=0}^\infty \sum\limits_{j=1}^{\left|\Omega(l)\right|}\text{P}\left[ n_{i(t+\d t)}(t+\d t)=1\: |\: (n_i)^l_{i=0, j}(t), L(t)=l \right] \text{P}\left[(n_i)^l_{i=0, j}(t) \:|\: L(t)=l\right]\text{P}\left[L(t)=l\right].
\end{equation}
\end{widetext}
The first factor is the probability that site $i$ is occupied at time $t+\d t$ under the condition that the system was $l$ sites long at time $t$ and its state was $(n_i)^l_{i=0, j}$. The second factor gives the probability that the system was in state $(n_i)^l_{i=0, j}$ at time $t$ under the condition that its length was $l$ and the last term corresponds to the probability that the system was $l$ sites long. Every possible state at fixed length and any length could contribute, hence the sums.
The difficulty is that the length distribution $\text{P}(L(t)=l)$ itself again depends on the occupancy numbers $\left\{n_i\right\}$, in particular on the TL tip occupancy $n_l^T$:
\begin{align}\label{eq:exactlengthchangeEq}
	\partial_t \text{P}(L=l) =& \delta n_{l+1}^T\text{P}(L=l+1) \\  &+\gamma \text{P}(L=l-1) - (\delta n_l^T+\gamma) P(L=l),\nonumber
\end{align} 
where the first two terms describe the probability gain due to a shrinkage or growth event of a longer or shorter length, respectively, while the last term represents the corresponding probability loss.

To tackle this problem analytically, we assume that the length changing dynamics happens at a far longer time scale than the particle hopping. Thus both dynamics can be decoupled. We refer to this simplification as \emph{adiabatic assumption}. It is untenable for large growth rates, as confirmed in the simulations, but suitable for small growth rates. In this regime the assumption implies that we can take the particle densities to adapt instantaneously to the current length and correspondingly that we can replace the (changing) length by a constant length when describing the particle occupancy dynamics. Thus, for the occupancy number dynamics, Eq.~(\ref{eq:exactoccupationnumberdynamics}), we choose the, by this assumption constant, length to equal the average lattice length. Mathematically this can be expressed by setting $P\p{L(t)=l}\propto\delta(l,\avg{L})$ where $\delta(i,j)$ is the Kronecker delta. 
On the other hand, in Eq.~(\ref{eq:exactlengthchangeEq}) for the length changing dynamics the actual occupancy $n_l^T$ can be replaced by its time average. The time average is equivalent to the ensemble average, $\rho_l$ at length $l$, that is the average tip occupancy at length $l$ (in contrast to the average occupancy at site $l$ for arbitrary length or the one for average length $\avg{L}$). We find

\begin{align}
	&\text{P}\p{n_{i(t+\d t)}(t+\d t)=1}= \\
	&\sum\limits_{j=1}^{\left|\Omega(\avg{L})\right|}\text{P}\left[n_{i(t+\d t)}(t+\d t)=1\: |\: (n_i)^{\avg{L}}_{i=0, j}(t)\right]\text{P}\left[(n_i)^{\avg{L}}_{i=0, j}(t)\right],\nonumber
\end{align}

and 
\begin{align}\label{eq: length changing}
\partial_t \text{P}(L=l) =& \delta \rho_{l+1}\text{P}(L=l+1)+\gamma \text{P}(L=l-1)\\ &- (\delta \rho_l+\gamma) P(L=l).\nonumber
\end{align}

So, applying the adiabatic assumption, we can decouple the occupancy number and length dynamics and proceed.

From now on, we will furthermore restrict ourselves to the \emph{stationary state} of the system, 
\begin{equation*}
\partial_t \avg{n_i}\overset{!}{=} 0
\end{equation*}
and $L\equiv \avg{L}$.
The next approximation to solve the coupled set of occupancy equations is to eliminate the correlations between occupancies at different sites by using the \emph{mean-field approximation}
\begin{equation*}
\avg{n_i n_j} \approx \avg{n_i}\avg{n_j}\equiv \rho_i\rho_j.
\end{equation*}
The equations for the occupancy dynamics at any site are then given by
\begin{align*}
0=\partial_t \rho_0 &= -\nu \rho_0 \p{1-\rho_1}- \omega \p{\rho_0-\mu_0 + \rho_0\mu_0},\\
0=\partial_t \rho_i &= \nu \p{\rho_{i-1} \p{1-\rho_{i}}- \rho_i\p{1-\rho_{i+1}}}\\ &- \omega \p{\rho_i - \mu_i + \rho_i\mu_i},\\
0=\partial_t\rho_L &= -\gamma \rho_L + \delta \rho_L\p{ \rho_{L-1}-1} + \nu \rho_{L-1} \p{1- \rho_L},\\
0=\partial_t \mu_0 &= \alpha - \epsilon \p{2 \mu_0- \mu_1} + \omega\p{\rho_0 + \rho_0\mu_0 - \mu_0},\\
0=\partial_t \mu_i &= \epsilon \p{\mu_{i+1}+ \mu_{i-1}- 2\mu_i}+\omega \p{\rho_i+\rho_i \mu_i -\mu_i},\\
0=\partial_t \mu_{L} &= -\gamma\mu_L + \delta \rho_L \p{1+\mu_{L-1}}+\epsilon \p{\mu_{L-1}-\mu_L},
\end{align*}
with $i$ denoting any bulk site.
The corresponding flux balances are
	\begin{align}
	\text{(TL)}\quad 0 &= + J_{0}^{D \to T} 
	+ J_1^T,  \\
	\text{(DL)}\quad 0 &= - J_0^{D \to T} 
	+ \epsilon (\eta_{1}-\eta_0) - \epsilon\eta_0 + \alpha ,\\
\text{(TL)}\quad 0 &= + J_i^{D \to T} 
+ (J_i^T - J_{i+1}^T)\, ,  \label{eq:currentbulkTL} \\
\text{(DL)}\quad 0 
&= - J_i^{D \to T}
+ D_i \, \label{eq:currentbulkDL}, 
\end{align}
for the first site and the bulk respectively.

Moreover, although the lattice is growing and shrinking, its length $L$ is typically 100 up to 1000 times larger than the remaining parameters and densities. Thus it is justified to consider the limit where the lattice spacing $\xi$ tends to zero when the total length of the system is rescaled to 1.

The second step is thus to apply the \emph{continuum limit} by replacing the lattice by a smooth interval $\left[0,1\right]$. Note that this is different to the choice in Sec.~\ref{sec:mathematical_analysis}, where the interval is set to $\left[0,L\right]$ in order to keep the notation cleaner. Here we want to clearly see the orders of the following Taylor expansion. We further define the occupancy density (also named $\rho$) to be the smooth function satisfying $\rho \p{\xi i/L}= \rho_{i}$ with $i = 1,\dots, L-1$ denoting the lattice site index. We set $\xi=1$ in order to rescale the system size to 1). $\rho$ can then be Taylor-expanded in the limit $1/L \rightarrow 0$:
\begin{equation*}
\rho\p{x\pm \frac{1}{L}} = \rho(x) \pm \frac{1}{L}\partial_x\rho(x) + \frac{1}{2L^2}\partial^2_x \rho(x) + \mathcal{O}\p{\frac{1}{L^3}}.
\end{equation*}
For the currents this expansion gives 
	\begin{align}
J^T(x) &= \left[\rho(x)-\frac{\partial_x}{L} \rho(x)\right]\left[1-\rho(x)\right]  \\
J^{D\rightarrow T}(x)&= \omega \left[1-\rho(x)\right]\eta(x)-\omega \rho(x)\\
D(x) &= \epsilon \frac{\partial_x^2}{L^2}\eta (x)\, .
\end{align}
Moreover, we focus on the \emph{mesoscopic limit} \cite{Parmeggiani2003, Parmeggiani2004} of $\omega$. This implies that $\omega=\Omega/L_0$, with $L_0$ denoting the initial length, is treated as order $1/L$. Consequently, $J^{D\rightarrow T}(x)$ has no 0th order contribution. 

As the DL is the only source of particles on the TL, we will at first solve the equation for the diffusive lane and use it to obtain the TL density profile.  We begin at the left boundary.
\begin{equation}
	\text{(DL)}\quad 0=-\epsilon \eta(0)+\alpha - J^{D\rightarrow T}(0)+\epsilon \frac{\partial_x}{L}\eta(0) \, ,
\end{equation}
thus, to 0th order in the lattice spacing we are left with $0=-\epsilon \eta(0)+\alpha$, concluding 
\begin{equation}
	\eta(0) = \frac{\alpha}{\epsilon} \,\label{eq:eta(0)} .
\end{equation}
For the TL we have
\begin{equation}
	0= \omega \left[1-\rho(0)\right]\eta(0)-\omega \rho(0)-\rho(0)\left[1-\rho(0)-\frac{\partial_x }{L}\rho(0)\right] \, ,
\end{equation}
implying $0=- \rho(0) (1-\rho(0)) + \mathcal{O}(1/L)$. This equation has two solutions, either the first site is always occupied or always empty. To lowest order, as we only treat the low density limit, the site has to be empty, i.e.\ $\rho(0)=0$. To first order, we obtain $0=\omega\eta(0)-\rho(0)$, thus
\begin{equation}
\rho(0)=\omega\eta(0) \, \label{eq:rho(0)}.
\end{equation}
The first order equation for the DL is 
\begin{equation}
	\omega \left[ \left(1-\rho(0) \right)\eta(0)-\rho(0) \right]=\epsilon \frac{1}{L}\partial_x \eta (0).
\end{equation} With $\rho(0)=0$ to 0th order, we obtain
\begin{equation}
	\frac{\partial_x}{L} \eta (0)=\lambda^{-2} \eta(0)\, \label{eq:deltarho(0)},
\end{equation}
with \begin{equation}
	\lambda\equiv \sqrt{\frac{\epsilon}{\omega}} \, .
\end{equation}
Having solved the boundary equations, we apply these results to solve the bulk equations. By adding the DL bulk dynamics equation corresponding to Eq.~\ref{eq:currentbulkDL}
\begin{equation}
	\text{(DL)}\quad 0=-J^{D\rightarrow T}+\epsilon\frac{\partial^2_x}{L^2}\eta(x)
\end{equation}
to the TL bulk dynamics equation derived from Eq.~\ref{eq:currentbulkTL}
\begin{equation}
\text{(TL)}\quad 0=J^{D\rightarrow T}(x)-\frac{\partial_x}{L}\rho(x)(2\rho(x)-1)\, ,
\end{equation}
we obtain the first order TASEP bulk equation $0=\partial_x\rho(x)(2\rho(x)-1)$. As we are in the low density limit, the solution $\rho(x)=1/2$, corresponding to the maximal current solution, can be ruled out, thus $\partial_x \rho(x)=0$. Using our results from the left boundary as initial values, we conclude that the constant density equals $\rho(x)=\rho(0)=0$ to first order. We conclude that the occupancy is constant and thus equals the occupancy at the first site. To first order, it has been determined to equal zero (Eq.~\ref{eq:rho(0)}), thus $\rho(x)=\rho(0)=0$. Inserting this result to the second order DL equation gives
\begin{equation}
	\epsilon \frac{\partial_x^2}{L^2} \eta (x)= \omega \eta(x) \, ,
\end{equation}
which is solved by 
\begin{equation}
	\eta(x)= A \sinh{\frac{x L}{\lambda}} + B \cosh{\frac{x L}{\lambda}} \, .
\end{equation}
Using our boundary conditions, $\eta(0)=\alpha/\epsilon$ (Eq.~\ref{eq:eta(0)}) and $(\partial_x/L) \eta(0)=\lambda^{-2}\eta(0)$ (Eq.~\ref{eq:deltarho(0)}), gives
\begin{equation}
	\eta(x)=\eta(0)\left( \frac{1}{\lambda}\sinh{\frac{x L}{\lambda}} +\cosh{\frac{x L}{\lambda}}\right)\, .
\end{equation}
We continue with the second order equations of the TL bulk, 
\begin{equation}
	0=\omega (1-\rho(x))\eta(x)-\omega\rho(x)+\frac{\partial_x}{L}\rho(x)(2\rho(x)-1) \, .
\end{equation}
Upon employing that the first order value of $\rho$ is zero (Eq.~\ref{eq:rho(0)}), we are left with
\begin{equation}
	0=\omega\eta (x)-\frac{\partial_x}{L}\rho(x) \, ,
\end{equation}
which is solved by 
\begin{equation}
	\rho(x)=\rho(0) + \frac{\alpha}{\lambda}\left(\frac{1}{\lambda}\cosh{\frac{x L}{\lambda}} + \sinh{\frac{x L}{\lambda}}\right) \, .
\end{equation}
In summary, we have found analytic expressions for the steady state TL and DL occupancy densities in the adiabatic limit.

\section{Oscillatory behavior}\label{appendix: Oscillatory behaviour}
In section \ref{sec:numerics} we have learned about the existence of a parameter regime where the length change exhibits oscillatory behavior. Following up on these investigations, we want to further discuss the methods used.
Moreover we want to examine the occupancy densities at the system tip when the system switches from growth to shrinkage.

As the time intervals between two events in the simulation are not uniformly spaced, we performed a Lomb Scargle analysis instead of a Fourier analysis \cite{Press1992} to determine the average oscillation frequency (see Fig.~\ref{fig:oscillationsFreqs}). The algorithm essentially fits a sine function to the data and checks which frequency matches the data best. We deduced the frequency with the smallest false alarm probability as well as the second and third best choice. For larger $\gamma$ values, the frequency decreases with increasing growth rate and reassuringly, the three best frequencies agree quite well. For small growth rates, the results should not be taken seriously, as there are also no visible oscillations in the time traces. 

Moreover, we determined the minima and maxima of a time-series of the length. This was done by cutting off the data of length for lengths larger than the initial length $L_0$ (which is 2-3 times larger than the minimum average length and smaller than the average length). Within each of the remaining intervals we determined the minimal length, while sorting out all minima that occurred very quickly after each other, i.e. in less time than a threshold $\Delta T$. This threshold excludes small fluctuations around $L=L_0$ and is chosen in a way to minimize artefacts of chopping off the length at $L_0$. We used $\Delta T = 800$. Note that our choice of $\Delta T$ does influence the frequency results as it restricts the maximal frequency. 
Between each two minima, we then determined the maxima. The respective averages and standard deviations for the maximally and minimally obtained system length during an oscillation period are plotted in Fig.~\ref{fig:oscillationsMinMax}. The maxima clearly increase with larger growth rates, whereas the minima remain rather constant. The latter further supports our intuition of a particle cluster at the DL tip, which equilibrates with the reservoir only when the system length is small enough for diffusion to be comparable to shrinkage. 
From the temporal distance of the minima, the oscillation frequency was deduced (see Fig.~\ref{fig:oscillationsFreqs}). The values agree with the result of the Lomb Scargle analysis mentioned before. 
As a third method to determine the oscillation frequency we extracted the frequency from the autocorrelation function (Eq.~\ref{eq:autocorr}). We searched for the first 2-4 maxima and minima of the autocorrelation function and averaged their distance. For smaller grow rates we had to reduce the number of maxima and minima, as the number of oscillations reduced from $>4$, to $2$ and even $1$ in the case of $\gamma = 0.005$. The extracted frequencies are also shown in Fig.~\ref{fig:oscillationsFreqs}.

\begin{figure}
	\centering
	\includegraphics[width=\linewidth]{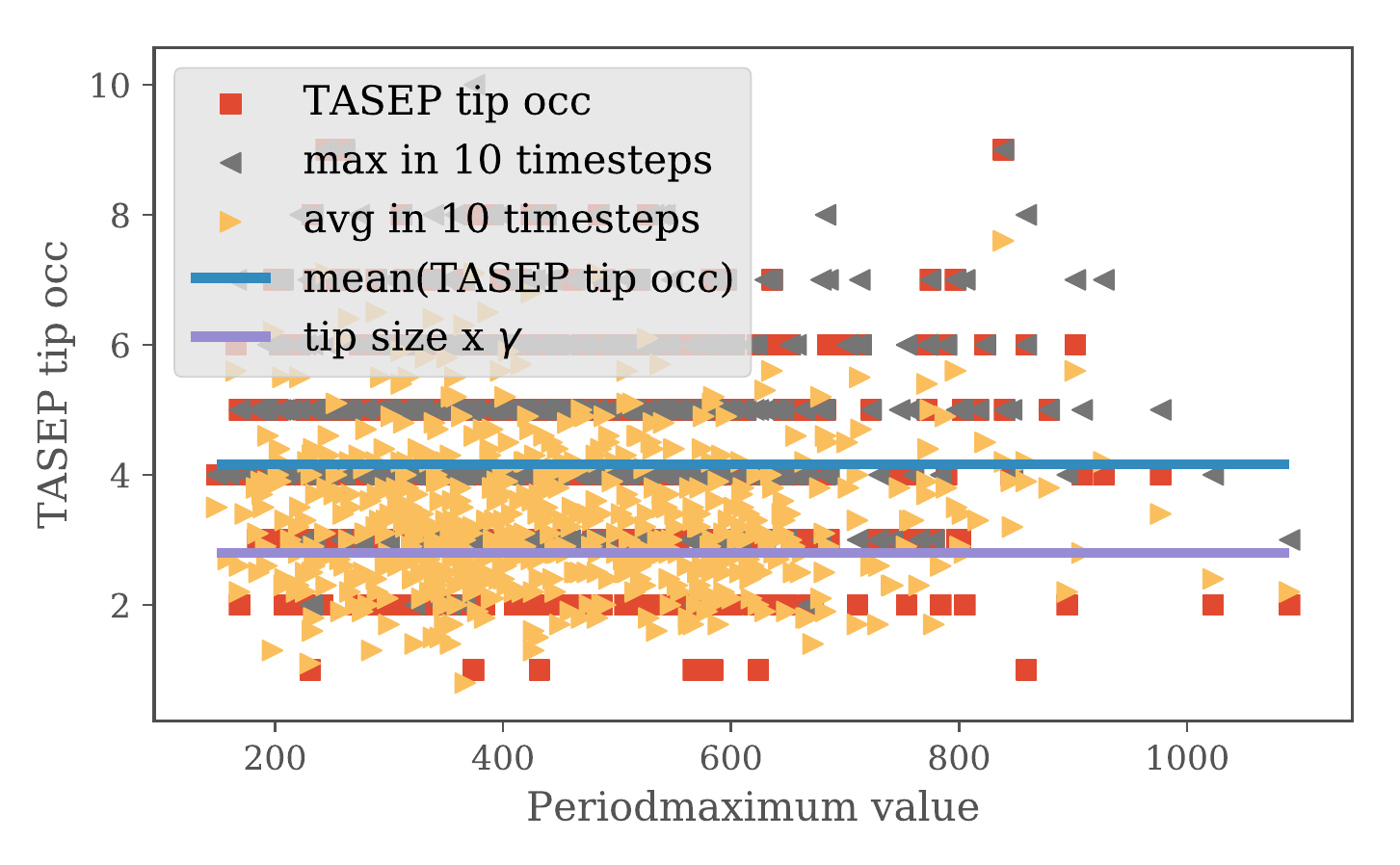}
	\caption{Occupancies of the TASEP lane tip neighborhood, \ie here the 20 last sites, at the time point or time interval when the length reaches its maximum during one oscillation period. They have been measured for several periods and are shown for a growth rate of 0.14. The occupancy at these time points (red squares) and their average (upper blue line) as well as the occupancy average (yellow right-pointing triangles) and maximum (gray left-pointing triangles) over a time period of ten events - five before and five after the maximum - is depicted. Moreover, the critical density for switching from growth to shrinkage (lower purple line) is shown.}
	\label{fig:suppturningpoint}
\end{figure}

Fig.~\ref{fig:suppturningpoint} shows the TASEP tip neighborhood (\ie 20 tip sites) occupancy at the oscillation maxima corresponding to the turning point between a phase of growth and a phase of shrinkage. The blue upper line represents the mean of the TASEP tip neighborhood occupancy (red squares). The results vary strongly, thus we checked related observables. But also the maximal tip neighborhood occupancy (gray left-pointing triangles) within ten timesteps - five before the maximum is reached and five thereafter - and the average (yellow right-pointing triangles) fluctuate. 
Nevertheless we see that nearly all measurements of the maximal tip neighborhood occupancy (gray left-pointing triangles) lie above the critical density (purple lower line), being $\gamma$ times the tip neighborhood size (here 20 sites) as the length change is given by $\partial_t L = \gamma - \delta \rho_+$ (as $\delta=1$). When we compare the critical density to the mean density for a time interval covering more than one oscillation period, and not just at the time points where the amplitude is maximal, the values coincide.
It can further be noted that none of the observables of the turning point tip occupancy increases for larger amplitudes, i.e. system lengths (for a fixed growth rate). 
These observations further support our intuition that the length grows until a critical occupancy density at the tip (depending solely on the growth rate) has been reached, triggering the switch to the shrinking phase. As shown in the time trace plot, Fig.~\ref{fig:timeSeries}, in section \ref{sec:numerics}, the total occupancy density follows a sawtooth-like trajectory. This is due to a constant influx from the reservoir during growth phase and most of the shrinkage phase. Only at the end of the shrinkage phase the cluster at the tip communicates with the reservoir and is quickly emptied.

\section{Detailed calculations for the effective theory} \label{appendix:effective}

In this section we elaborate on the mathematical details for the effective theory. First, in Section~\ref{appendix:effective_in_rate} we determine the effective in-rate from the reservoir onto the TL. In order to estimate the density in the bulk, we infer the length of the in-region, see Section~\ref{appendix:effective_in_region}. Finally, we deduce the recursion relation for the tip density in Section~\ref{appendix:effective_recursion_relation}, and the minimal length in Section~\ref{appendix:effective_minimal_length}.

\subsection{Effective in-rate $\alpha_{\mathrm{eff}}$} \label{appendix:effective_in_rate}

In this subsection, we comment on how we determine the effective rate at which particles enter the system and then attach to the TL. This rate will not equal the ``bare" in-rate $\alpha$ as particles can also leave the system before attaching. What is the probability that a particle that enters from the reservoir leaves the system again, before attaching to the TL? To answer this question, let us consider a situation where the length of the system is considerably larger than the length of a typical journey of a particle on the DL before attaching to the TL, the latter of which we estimate as $\lambda \pm \lambda$ (see later). Here, $\lambda = \sqrt{\epsilon/\omega}$ is the characteristic length scale of the system. In this case of large length, the probability that a given particle that enters the DL leaves back into the reservoir before attaching to the TL, $\mathrm{Prob} (\mathrm{leaving})$, is given by:
\begin{align}
\mathrm{Prob} (\mathrm{leaving}) = \sum_{j=0}^{\infty} p q^j A^j \, ,
\label{eq:appendix_prob_leaving}
\end{align}
where $p = \tilde{\epsilon}/\left(\tilde{\epsilon}+\epsilon +\omega\right)$ is the probability that a particle exits from the first DL site back into the reservoir \footnote{We allow here that the rate at which a particle leaves into the reservoir, $\tilde{\epsilon}$, can be distinct from the usual diffusion rate $\epsilon$.}. The quantity $q = \epsilon/\left(\tilde{\epsilon}+\epsilon +\omega\right)$ is the probability that a particle proceeds to diffuse into the protrusion and $A$ is the probability that a particle that starts at the first site of the DL returns to the first site of the DL without attaching to the TL in between. 
Since returning to the first site of the DL without attaching to the TL in between can only happen after an even number of steps on the DL, this probability $A$ comprises the probabilities that the particle diffuses back to the first site of the DL in exactly $2j$ steps, $j \in \mathbb{N}$, without attaching to the TL. The latter are given by the product of the probability that a symmetric random walker returns back to its starting point after exactly $2 j$ steps and the probability that the particle stays on the DL in each step.
Taken together we determine $A$ as:
\begin{widetext}
\begin{align*}
A&= \mathrm{Prob} (\mathrm{return \ to \ site \ 1 \ w/o \ attaching \ to \ TL}) = \\
&= \sum_{j=1}^{\infty} \mathrm{Prob} (\mathrm{return \ to \ site \ 1 \ in \ exactly \ 2j \ steps \ w/o \ attaching \ to \ TL}) =\\
&=\sum_{j=1}^{\infty} \mathrm{Prob} (\mathrm{return \ to \ site \ 1 \ in \ exactly \ 2j \ steps \ } | \mathrm{ \  not \ attaching \ to \ TL \ during \ the \ } 2j \mathrm{ \ steps}) \times \\
& \hspace{20pt} \times \mathrm{Prob} (\mathrm{not \ attaching \ to \ TL \ during \ the \ } 2j \mathrm{ \ steps}) =\\
&= \sum_{j=1}^{\infty} f_{2j} \left( \frac{2 \epsilon}{2 \epsilon + \omega}\right)^{2 j-1} = \frac{2 \epsilon +\omega}{2 \epsilon} \left( 1-\frac{\sqrt{\omega (4 \epsilon+ \omega)}}{2 \epsilon + \omega}\right)
\end{align*}
\end{widetext}
where $f_{2j} = \binom{2 j}{j} /\left[ (2j-1) 2^{2j} \right]$ is the probability that a symmetric 1D random walker returns to its starting point for the first time in exactly $2j$ steps. Note that the probability that the particle does not attach during the $2 j$ steps only has $2 j\!-\!1$ terms $2 \epsilon/\left(2 \epsilon + \omega\right)$ as the first step into the protrusion is already accounted for by the probability $q$ in $\mathrm{Prob} (\mathrm{leaving})$, Eq.~(\ref{eq:appendix_prob_leaving}). Combining the result for $A$ with the explicit formulas for $p$ and $q$ we find
\begin{align*}
\mathrm{Prob} (\mathrm{leaving}) = 1-\frac{\omega +\sqrt{\omega (4 \epsilon +\omega)}}{\omega +\sqrt{\omega (4 \epsilon +\omega)}+2 \tilde{\epsilon}} \, .
\end{align*}
We approximate this formula for small $\omega$ by Taylor expanding up to first order in $\omega$:
\begin{align*}
\mathrm{Prob} (\mathrm{leaving}) =  1- \frac{\sqrt{\omega \epsilon}}{\tilde{\epsilon}} + \frac{2 \epsilon - \tilde{\epsilon}}{2 \tilde{\epsilon}^2} \omega + \mathcal{O} (\omega^{3/2}) \, .
\end{align*}
The effective in-rate is given by the ``bare" in-rate $\alpha$ weighted by the probability that a particle that enters from the reservoir attaches to the TL. The latter probability is just $1 \!-\! \mathrm{Prob} (\mathrm{leaving})$.
This implies that the effective in-rate is given by 
\begin{align}
\alpha_{\mathrm{eff}} = \alpha \ \frac{\omega +\sqrt{\omega (4 \epsilon +\omega)}}{\omega +\sqrt{\omega (4 \epsilon +\omega)}+2 \tilde{\epsilon}} \approx \frac{\alpha }{\tilde{\epsilon}} \sqrt{\omega \epsilon}
\label{eq:effalpha_supp}
\end{align}
to lowest order in $\omega$. 

\subsection{Length of the in-region and density in the bulk} \label{appendix:effective_in_region}

To continue we now estimate the length of the ``in-region" $l_I$ since - due to the attachment and detachment - the density at the end of the in-region depends on the length.  For this, we look at a symmetric random walk with reflecting boundary at $x=0$ as we want to find out the typical journey of a particle on the DL that is eventually attaching to the TL (and thus not leaving the system again). Using the initial condition $p(x,t=0) = \delta (x)$, the probability distribution of such a process is given by $p(x,t) = e^ {-x^2/ (4 \epsilon t)}/\sqrt{\pi \epsilon t}$ for $x \geq 0$ where $\epsilon$ is the diffusion constant (lattice spacing 1). 
To determine the average distance a particle travels on the DL before attaching to the TL, $\langle x \rangle_{\mathrm{attach}}$,  and its standard deviation, $\sigma_{\mathrm{attach}}$, we need to take two processes into account. First, we need to find out how the time at which the particle attaches to the TL is distributed and, second, how far a particle travels until a certain time point. Using that the attachment process is a Poisson process of rate $\omega$, where the time until an attachment event happens has the probability distribution $f(t) = \mathrm{Prob} (T=t) = \omega e^{- \omega t}$, we calculate the mean and variance as
$\langle x \rangle_{\mathrm{attach}}  = \int_0^{\infty} \mathrm{d} t\ f(t) \langle x (t) \rangle$ and $\sigma_{\mathrm{attach}}^2 = \int_0^{\infty} \mathrm{d} t\ f(t) \left( \langle x^2 (t) \rangle \!-\! \langle x (t) \rangle^2 \right)$. Here, $\langle x (t) \rangle$ and $\langle x^2 (t) \rangle \!-\! \langle x (t) \rangle^2$ are the mean and variance of the travelled distance of the symmetric random walk until time $t$. For those quantities we find $\langle x (t) \rangle = 2 \sqrt{\epsilon t/\pi}$ and $\langle x (t)^2 \rangle = 2 \epsilon t$ from the above probability distribution $p(x,t)$. As a result, $\langle x \rangle_{\mathrm{attach}}  = \sqrt{\epsilon/\omega} = \lambda$ and $\sigma_{\mathrm{attach}} = \lambda = \langle x \rangle_{\mathrm{attach}}$.
Therefore, $\langle x \rangle_{\mathrm{attach}} \pm \sigma_{\mathrm{attach}} = \lambda \pm \lambda$, which means that the standard deviation is the same as the mean. We will thus approximate the length of the in-region as twice the average distance a particle travels before attaching to the TL: 
\begin{align}
l_I \approx 2 \lambda.
\end{align}

With this relation, we now determine the density profile in the in-region which we assume to equilibrate quickly on the time scale of the oscillations.  Let us denote by $\tilde{\rho}_i$ the density at site $i$ from the base. Then, we approximate the time evolution of the density at site $i=1, \ldots, L_I$ as
\begin{align*}
0=\partial_t \tilde{\rho}_i &\approx \tilde{\rho}_{i-1} - \tilde{\rho}_i + \frac{\alpha_{\mathrm{eff}}}{L_I} - \omega \tilde{\rho}_i
\end{align*}
in the low-density limit with the boundary condition $\tilde{\rho}_0 \approx 0$. Here, $L_I = \mathrm{Round} (l_I)$ is the integer length of the in-region within which we assume homogeneous attachment (at rate $\alpha_{\mathrm{eff}}/L_I$ per site). Performing a continuum limit $i \rightarrow x$, $x\in [0, l_I]$, and considering only the first derivative with respect to $x$, we have:
\begin{align*}
0=-\partial_x \tilde{\rho}(x) +\frac{\alpha_{\mathrm{eff}}}{l_I} -\omega \tilde{\rho}(x)
\end{align*}
with the solution $\tilde{\rho} (x) = \alpha_{\mathrm{eff}}(1-e^{- \omega x})/ (l_I \omega)$. So, in particular, we obtain for the particle density at the end of the in-region
\begin{align*}
\tilde{\rho} (l_I) = \frac{\alpha}{2 \tilde{\epsilon}} \left( 1-e^{- 2 \sqrt{\epsilon \omega}}\right) \, ,
\end{align*}
where we combined all the above results. 

\subsection{Recursion relation for the tip density} \label{appendix:effective_recursion_relation}

As mentioned in Section~\ref{sec: Effective theory}, for the analysis of the tip neighborhood and the tip, we go to a different reference frame, namely starting at the tip and reaching into the tip neighborhood, co-moving with the tip. By $\rho_0 \equiv \rho^{+}$ we denote the density at the tip and by $\rho_i$ the density at the $i$-th site from the tip. Since we defined the tip neighborhood to be the region where motors that have previously detached at the tip reattach to the TL, we assume that the tip neighborhood has the same length as the in-region $l_T= l_I= 2 \lambda$ as for both the average distance before attaching to the TASEP becomes essential. 

Note that we ignore the influence of the growth and shrinkage dynamics on the average distance before attaching. This, however, should be legitimate in our parameter regime: Assume that we look at a symmetric random walk on a lattice with one reflecting boundary at the left end where also new particles are injected. At each site, the particles can leave the system at rate $\omega$, and the reflecting boundary moves at rate $v>0$ to the right (or, in case of $v<0$, at rate $-v$ to the left). Then, in the co-moving frame (moving with the reflecting boundary), the steady-state profile is proportional to $e^ {- x/\bar{\lambda}}$ with the length scale $\bar{\lambda} = 2 \epsilon/ \left( v+\sqrt{v^2 + 4 \epsilon \omega} \right)$ which corresponds to the average travelled distance before leaving the system via $\omega$. In our case, the velocity is not constant but if we assume that the velocity is homogeneously distributed in $[ -\gamma, \gamma ]$, we get an average length scale which is very close to $\lambda$ for our choice of parameters.

Let us now go back to the densities in the tip neighborhood and right at the tip. By taking into account the reattachment of motors that have detached at the tip an (average) time $1/\omega$ before, the growth and shrinkage dynamics, and the usual hopping, we find for the time evolution of the density $\rho_i$ in the low-density limit:
\begin{align}
\partial_t \rho_i {=} \rho_{i{+}1} {-} \rho_i {+} \frac{\delta \rho^{+}_{\mathrm{before}}}{l_T} {+} \gamma (\rho_{i{-}1}-\rho_i) {+} \delta \rho^{+} (\rho_{i{+}1}-\rho_i),
\label{eq:tn1}
\end{align}
and for the tip density $\rho_0=\rho^{+}$
\begin{align}
\partial_t \rho^{+} {=} \rho_1 {-} \gamma \rho^{+} {-} \delta \rho^{+} (1{-}\rho_1).
\label{eq:tn2}
\end{align}
Note, however, that in the last equation for the tip density we take exclusion into account explicitly by assuming that the occupancy at the tip is exactly 1 in case of a shrinkage event (last term). If exclusion was lifted, there could be more than one particle at the tip and several particles would then be simultaneously released into the cytosol in case of a shrinkage event.\\
For the time evolution of $\rho_i$, Eq.~(\ref{eq:tn1}), we assume that the particles that have previously detached at the tip (at rate $\delta \rho^{+}_{\mathrm{before}}$, corresponding to a previous tip density $\rho^{+}_{\mathrm{before}}$), homogeneously reattach to the TL in the tip neighborhood. 
To proceed we now make the following ansatz: We assume that for a tip density $\rho^{+}_{\mathrm{before}}$ at time $t$ we can determine the tip density at time $t' = t + 1/\omega + \lambda = t + \Delta$ by solving Eqs.~(\ref{eq:tn1}-\ref{eq:tn2}) for $\rho^{+}$ in the steady-state. The idea behind this is that a particle that has detached at the tip needs on average $1/\omega$ to reattach to the TL, and then has to walk on average $\lambda$ sites to get back to the tip (we measure time in units of $\nu \equiv 1$, and length in units of the lattice spacing $a \equiv 1$). Using the continuum approximation in Eq.~(\ref{eq:tn1}) and considering only zero- and first-order terms, we find for the density in the tip neighborhood
\begin{align*}
\rho (x) = \tilde{\rho}(l_I) + \frac{\delta \rho^{+}_{\mathrm{before}}}{l_T} \frac{1}{1+\delta \rho^{+} - \gamma} (l_T+1-x) \, ,
\end{align*}
where we used the boundary condition $\rho(l_T +1) = \tilde{\rho}(l_I)$. As a result, the density at the 
site next to the tip is given by
\begin{align}
\rho_1 = \rho(1) = \tilde{\rho} (l_I) + \frac{\delta \rho^{+}_{\mathrm{before}}}{1+\delta \rho^{+} - \gamma}.
\label{eq:appendix_rho1}
\end{align}
Combining this with Eq.~(\ref{eq:tn2}) we solve for $\rho_1$ and find an equation for the tip density $\rho^{+}$ in terms of the previous tip density $\rho^{+}_{\mathrm{before}}$:
\begin{align*}
( \delta + \gamma ) \rho^{+} ( 1 - \gamma + \delta \rho^{+} ) = \\
= ( 1 + \delta \rho^{+} )  \left[ ( 1 - \gamma {+} \delta \rho^{+} ) \tilde{\rho}(l_I)  + \delta \rho^{+}_{\mathrm{before}} \right] .
\end{align*}
Bearing in mind that the tip density should be positive, this equation is solved by
\begin{widetext}
\begin{align*}
\rho^{+} &= \left[\delta^2 \rho^{+}_{\mathrm{before}} - A+ \sqrt{\left(\delta^2 \rho^{+}_{\mathrm{before}} -A \right)^2 + 2 B \left( \delta \rho^{+}_{\mathrm{before}} +C \right)} \right]/B
\end{align*}
\end{widetext}
where $\tilde{\rho} (l_I) = \alpha \left( 1-e^{-2 \sqrt{\epsilon \omega}}\right)/ \left( 2 \tilde{\epsilon} \right)$ (see above) and
\begin{align*}
A&= \delta \left(1 - \gamma \right) + \gamma \left(1-\gamma \right) - \delta \tilde{\rho} (l_I) \left(2-\gamma \right), \\
B&=2 \delta \left[\delta (1-\tilde{\rho}(l_I)) + \gamma \right],  \\
C&=(1-\gamma) \tilde{\rho} (l_I).
\end{align*}
So, this equation relates the previous tip density $\rho^{+}_{\mathrm{before}}$ at time $t$ to the tip density $\rho^{+}$ at time $t+ \Delta$. Iterating this procedure, we find a recursion relation for the tip densities $\rho^{+}_{\mathrm{it}}$ at times $t_{\mathrm{it}} = \mathrm{it} \times \Delta$: 
\begin{align}
\rho^{+}_{\mathrm{it}} & {=} \left[\delta^2 \rho^{+}_{\mathrm{it} {-} 1} {-} A {+} \sqrt{\left(\delta^2 \rho^{+}_{\mathrm{it} {-} 1} {-} A \right)^2 {+} 2 B \left( \delta \rho^{+}_{\mathrm{it} {-} 1} {+} C \right)} \right] {/} B.
\label{eq:soltip}
\end{align}

\subsection{Minimal length} \label{appendix:effective_minimal_length}

So far, we have considered the situation where the length of the system is much longer than the average distance a particle typically travels on the DL. However - if the system is too small - the particles do not reattach to the TL as they leave the system too quickly. As a result, most of the particles will have left the system before the system is shrunk to zero, and the system will regrow from a minimal length larger than zero. To estimate this minimal length, let us consider a 1D system of length $l$ with injection of particles at rate $r$ at site $l$, symmetric diffusion at rate $\epsilon$ within the system, outflux of particles at rate $\omega$ everywhere, and an additional outflux of particles at rate $\tilde{\epsilon}$ at site $0$. In the steady-state and with a continuum approximation we thus have
\begin{align*}
0&=\partial_t p(x,t) = \epsilon \partial_x^2 p(x,t) - \omega p(x,t) \, ,\\
0&=\partial_t p(0,t) = \epsilon \partial_x p(0,t) - (\omega+\tilde{\epsilon}) p(0,t) \, , \\
0&=\partial_t p(l,t) = -\epsilon \partial_x p(l,t)- \omega p(l,t) + r \, .
\end{align*}
Those equations are solved by
\begin{widetext}
\begin{align*}
p(x) =  \frac{r e^{\frac{\left(l-x \right)}{\lambda}}}{\omega} \frac{ \varphi \lambda \left( 1+e^{\frac{2 x}{\lambda}} \right) + \left( \varphi + \lambda^2 \right) \left( -1+e^{\frac{2 x}{\lambda}} \right)}{\lambda \left(2 \varphi +\lambda^2 \right) \left( 1+e^{\frac{2 l}{\lambda}} \right) + \left( \varphi + \left(\varphi+1 \right) \lambda^2 \right) \left( -1+e^{\frac{2 l}{\lambda}} \right)} ,
\end{align*}
\end{widetext}
where  $\varphi = \epsilon/\tilde{\epsilon}$ is the ratio between the diffusion and the exit rate. So, we determine the (steady-state) probability that a particle that enters the system at site $l$ (the tip) exits it via the rate $\tilde{\epsilon}$ (back into the reservoir) and not via $\omega$ (attaching to the TL) as $p_{\mathrm{exit}} = p(0) \tilde{\epsilon}/r$, which yields
\begin{align*}
p_{\mathrm{exit}}
=  \frac{ 2 \lambda^3 e^{\frac{l}{\lambda}} }{\lambda \left(2 \varphi {+} \lambda^2 \right) \left( 1 {+} e^{\frac{2 l}{\lambda}} \right) {+} \left( \varphi {+} \left(\varphi {+}1 \right) \lambda^2 \right) \left( e^{\frac{2 l}{\lambda}} {-} 1 \right)}.
\end{align*}
As a result, the reattachment probability for a particle detaching at the tip at length $l$ is approximated as
\begin{align}
p_{\mathrm{reattach}} (l) \approx 1 - F e^{-\frac{l}{\lambda}}
\label{eq:appendix_reattachment}
\end{align}
for $l \gg \lambda$. Here, $F = 2 \lambda^3/ \left(\varphi + 2 \varphi \lambda + \left(\varphi + 1 \right) \lambda^2 + \lambda^3 \right)$.
This means that the probability that a particle has not yet left the system after a series of lengths $\{ l_{\mathrm{it}}\}_{ \mathrm{it} =1, \ldots, n}$ is given by
\begin{align*}
p_{\mathrm{survival}} \left(\{l_{\mathrm{it}}\}_{\mathrm{it}=1, \ldots, n} \right) = \prod_{ \mathrm{it} =1}^n p_{\mathrm{reattach}} (l_{\mathrm{it}})
\end{align*}
or, equivalently,
\begin{align*}
\ln \left[ p_{\mathrm{survival}} \left(\{l_{\mathrm{it}} \}_{ \mathrm{it} =1, \ldots, n} \right) \right]= \sum_{ \mathrm{it}=1}^n \ln \left[ p_{\mathrm{reattach}} (l_{\mathrm{it}}) \right].
\end{align*}
Assuming that the system shrinks at constant velocity $v$: $l(t) = l_0- vt$, and taking into account that in our effective system each length is realized for time $\Delta$ (during this time a particle that has detached potentially reattaches and walks back to the tip), we identify the length dynamics until time t with $\{ l_0, l_0 - v \Delta, \ldots, l_0 - v \Delta (t/\Delta) \}$. Approximating the sum as an integral, we then deduce the ``survival" probability until time $t$ as
\begin{align}
\ln \left[ p_{\mathrm{survival}} \left(t \right) \right] &\approx  \sum_{k=0}^{\frac{t}{\Delta}} \ln \left[ p_{\mathrm{reattach}} (l_0 - v \Delta k) \right] \approx \nonumber \\
&\approx \int_0^{\frac{t}{\Delta}} \mathrm{d} k \ \ln \left[ p_{\mathrm{reattach}} (l_0 - v \Delta k) \right] = \nonumber \\
&= \frac{1}{\Delta} \int_0^t \mathrm{d} t' \ \ln \left[ p_{\mathrm{reattach}} (l(t')) \right] = \nonumber \\
&= \frac{1}{v \Delta} \int_{l(t)}^{l_0} \mathrm{d} l \ \ln \left[ p_{\mathrm{reattach}} (l) \right] \approx \nonumber \\
&\approx \frac{1}{v \Delta} \int_{l(t)}^{\infty} \mathrm{d} l \ \ln \left[ p_{\mathrm{reattach}} (l) \right],
\label{eq:survival}
\end{align}
where we used the coordinate transformations $t'= k \Delta$, $l(t') = l_0 - v t'$, and approximated the maximal length of the system $l_0$ by $\infty$ as for maximal length the reattachment probability should be close to 1. Approximating the logarithm as $\ln \left[ p_{\mathrm{reattach}} (l) \right] \approx -1+ p_{\mathrm{reattach}}$ for $p_{\mathrm{reattach}} \geq 0.9$ we thus find
\begin{align*}
\ln \left[ p_{\mathrm{survival}} \left(t \right) \right] \approx  -\frac{1}{v \Delta} \int_{l(t)}^{\infty} \mathrm{d} l \ \left[ 1- p_{\mathrm{reattach}} (l) \right].
\end{align*}
Finally, using Eq.~(\ref{eq:appendix_reattachment}) for $1- p_{\mathrm{reattach}} (l)$ for $l \gg \lambda$ we get
\begin{align*}
\ln \left[ p_{\mathrm{survival}} \left(t \right) \right] \approx -\frac{\lambda F}{v \Delta} e^{-\frac{l(t)}{\lambda}}.
\end{align*}
Defining the average minimal length as the length where the probability that a particle that was in the system at maximal length has left the system is just 0.5 we find $
l_{\mathrm{min}} \approx \lambda \ln \left[ \lambda F / \left( v \Delta \ln (2) \right) \right]$.
For the (constant) velocity we make a very crude approximation, namely $v \approx \gamma/2$, and we have
\begin{align*}
l_{\mathrm{min}} \approx \lambda \ln \left[\frac{ 2 \lambda F}{\gamma \Delta \ln (2)} \right].
\end{align*}
\cleardoublepage

\end{document}